\documentclass[preprint,prd,nofootinbib,showpacs]{revtex4}
\usepackage{graphicx}
\usepackage{epsfig}
\usepackage{bm}
\begin{document}   

\title{\large\bf $\Lambda_b \to (\Lambda_c,\,p)\,\tau\,\nu$ decays within standard model and beyond}
\author{Rupak~Dutta${}$}
\email{rupak@phy.nits.ac.in}   
\affiliation{
National Institute of Technology Silchar, Silchar 788010, India\\
}

\begin{abstract}
Deviations from the standard model prediction have been observed in several leptonic and semileptonic $B$ meson decays to $\tau\nu$ final states mediated via $b \to u$ and $b \to c$ charged current interactions. The measured value of ratio of branching ratios $R_{\pi}^l$ of $B^- \to \tau^-\,\nu_{\tau}$ to $B^0 \to \pi^+\,\l^-\,\nu$ decays, where $l = (e,\,\mu)$, is larger than the standard model prediction by more than a factor of two. Similarly,  a combined excess of $3.9\sigma$ from the standard model expectation has been reported by HFAG for the values of $R_D$ and $R_{D^{\ast}}$, where $R_{D,\,D^{\ast}}$ represents the ratio of branching ratios of $B \to (D,\,D^{\ast})\,\tau\nu$ to corresponding $B \to (D,\,D^{\ast})\,l\nu$ decays, respectively. Very recently, hint of lepton flavor violation has been observed in the ratio of branching fractions of $B \to K\,e^+\,e^-$ to $B \to K\,\mu^+\,\mu^-$ decays as well. In this context, 
we employ an effective Lagrangian approach to study the decay branching fractions and the ratio of branching fractions of $\Lambda_b \to \Lambda_c\,l\,\nu$ and $\Lambda_b \to p\,l\,\nu$ decays within the standard model and beyond. We constrain the new physics parameter space using the existing experimental data on $R_D$, $R_{D^{\ast}}$, and $R_{\pi}^l$. We give predictions for various observables in the context of various new physics scenarios.
\end{abstract}


\maketitle

\section{Introduction}
\label{int}
Hint of lepton flavor violation has been observed in various leptonic and semileptonic $B$ decays. Recently, LHCb collaboration~\cite{Aaij:2014ora} has  measured the ratio of branching fractions of $B \to K\,e^+\,e^-$ to $B \to K\,\mu^+\mu^-$ decays to be $R_K^{\mu\,e} = 0.745^{+0.090}_{-0.074}$ in the dilepton invariant mass squared range $(1< q^2<6)\,{\rm GeV^2}$. It differs from the standard model~(SM) expectation at $2.6\sigma$ significance level. Similar tensions between theory and experiment have been observed in $B \to \tau\,\nu$ and $B \to (D,\,D^{\ast})\,\tau\,\nu$ decays mediated via $b \to u$ and $b \to c$ charged current interactions as well~\cite{Lees:2012xj, Lees:2013uzd, Huschle:2015rga, Aaij:2015yra}. A combined excess of $3.9\sigma$ from the SM prediction have been reported by HFAG  on $R_D$ and $R_{D^{\ast}}$, where
\begin{eqnarray}
\label{rdrds}
&&R_D = \frac{\mathcal B(\bar{B}^0 \to D\,\tau\,\nu)}{\mathcal B(\bar{B}^0 \to D\,l\,\nu)} = 0.391\pm 0.041\pm 0.028\,, \nonumber \\
&&
R_{D^{\ast}} = \frac{\mathcal B(\bar{B}^0 \to D^{\ast}\,\tau\,\nu)}{\mathcal B(\bar{B}^0 \to D^{\ast}\,l\,\nu)} = 0.322 \pm 0.018 \pm 0.012\,.
\end{eqnarray}
Again, there is a discrepancy of more than $2\sigma$ with the SM expectation in the measured ratio $R_{\pi}^l = 0.73\pm 0.15$~\cite{Fajfer:2012jt},
\begin{eqnarray}
R_{\pi}^l = \frac{\tau_{B^0}}{\tau_{B^-}}\,\frac{\mathcal B(B^- \to \tau^-\,\nu_{\tau})}{\mathcal B(B^0 \to \pi^+\,l^-\,\nu_l)}
\end{eqnarray}
where $l$ represents either an electron or a muon, respectively.
The recent value of $\mathcal B(B^- \to \tau^-\,\nu_{\tau}) = (11.4\pm 2.2)\times 10^{-5}$~\cite{belle, babar, pdg} is slightly larger than the SM expectation~\cite{Charles, ckm, Bona}. Again, the most recent result of $\mathcal B(B^- \to \tau^-\,\nu_{\tau}) = (12.5\pm 2.8 \pm 2.7)\times 10^{-5}$ reported by Belle~\cite{Abdesselam:2014hkd} is consistent with their earlier result. Moreover, the measured value of $\mathcal B(B^0 \to \pi^+\,l^-\,\nu_l) = (14.6 \pm 0.7)\times 10^{-5}$~\cite{delAmoSanchez:2010af, Ha:2010rf, Asner:2010qj} is consistent with the SM prediction. The Belle experiment recently reported an upper limit on the total rate $\mathcal B(B^0 \to \pi^-\tau\nu) < 2.5\times 10^{-4}$~\cite{Hamer:2015jsa} which is close to the SM prediction~\cite{Du:2015tda}. A prediction on the ratio of branching ratios $R_{\pi}$ of $B \to \pi\tau\nu$ to the corresponding $B \to \pi\,l\,\nu$ decays has also been reported in Refs.~\cite{Du:2015tda, Dutta:2013qaa, Bernlochner:2015mya, Chen:2006nua, Khodjamirian:2011ub}. Several phenomenological work have been done in order to explain the discrepancies in $R_D$, $R_{D^{\ast}}$, $R_{\pi}^l$, and $R_K^{\mu\,e}$, see in particular Ref.~\cite{Fajfer:2012jt, Dutta:2013qaa, Krawczyk:1987zj, Chen:2006nua, Kalinowski:1990ba, Hou:1992sy, Tanaka:1994ay, Kamenik:2008tj, Nierste:2008qe, Tanaka:2010se, Fajfer:2012vx, Sakaki:2012ft, Datta:2012qk, Soffer:2014kxa, Bailey:2012jg, Becirevic:2012jf, Tanaka:2012nw, Freytsis:2015qca, Bhattacharya:2015ida, Celis:2012dk, Ko:2012sv, Crivellin:2012ye, Deshpande:2012rr, Sakaki:2013bfa, Greljo:2015mma, Dorsner:2013tla, Biancofiore:2013ki, Fan:2015kna, Hati:2015awg, Glashow:2014iga, Bhattacharya:2014wla}. Ratio of branching ratios such as $R_D$, $R_{D^{\ast}}$, $R_{\pi}^l$, $R_{\pi}$, and $R_K^{\mu\,e}$ are excellent observables to test for new physics~(NP) mainly because of two reasons. First, these ratio of branching ratios are independent of the CKM matrix element and hence the uncertainties associated with the CKM matrix elements do not enter into these ratios. Second, uncertainties associated with the hadronic form factors are also reduced while taking these ratios.

Precise determination of the CKM matrix elements $|V_{cb}|$ and $|V_{ub}|$ is interesting in itself as there are tensions between exclusive and inclusive determination of $|V_{cb}|$ and $|V_{ub}|$ from semileptonic $B$ decays. The typical relative accuracy is about $2\%$ for $|V_{cb}|$, however,  the precision on $|V_{ub}|$ is not better than $12\%$~\cite{Hicheur:2015oka}. The magnitude $|V_{ub}|$ can be measured via semileptonic $b \to u$ transition decays. The world average using the exclusive $b \to u$ transition decay channels $\bar{B}^0 \to \pi^+\,l\,\nu$ and $B^- \to \pi^0\,l\,\nu$ is $|V_{ub}| = (32.8 \pm 2.9)\times 10^{-4}$~\cite{Amhis:2014hma}. Very recently, LHCb has measured the ratio of partially integrated rates of baryonic $b \to u$ and $b \to c$ decays $\Lambda_b \to p\,\mu\nu$ and $\Lambda_b \to \Lambda_c\,\mu\nu$ and put constraint on the ratio of $|V_{ub}|$ and $|V_{cb}|$. Combining with the theoretical calculations and previously measured value of $|V_{cb}| = (39.5 \pm 0.8) \times 10^{-3}$~\cite{pdg}, the obtained value of $|V_{ub}| = (32.7\pm2.3)\times 10^{-4}$~\cite{Aaij:2015bfa,Fiore:2015cmx} is in good agreement with the exclusively measured world average. However, it disagrees with the inclusive measurement at a $3.5\sigma$ significance level. This is the first measurement of $|V_{ub}|$ using baryonic decay channels. The baryonic $\Lambda_b \to p\mu\nu$ decays mediated via $b \to u$ charge current interactions was not considered before as $\Lambda_b$ baryons are not produced in $e^+\,e^-$ $B$ factory. However, at LHC, production of $\Lambda_b$ baryon is remarkably high; around $20\%$ of the total $b$ hadrons produced~\cite{Aaij:2011jp, Aaij:2014jyk}.

$\Lambda_b \to \Lambda_c\,\tau\,\nu$ decay mode has been studied by various authors~\cite{Woloshyn:2014hka, Shivashankara:2015cta, Gutsche:2015mxa, Detmold:2015aaa}. In Ref.~\cite{Shivashankara:2015cta}, a prediction for the decay branching fractions and the ratio of branching fractions has been presented in the context of SM and various NP couplings. In Ref.~\cite{Gutsche:2015mxa}, a covariant confined quark model has been used to provide SM prediction on various observables such as total rate, the differential decay distribution, the longitudinal and transverse polarization of the daughter baryon $\Lambda_c$ and the $\tau$ lepton, and the lepton side forward backward asymmetries. Again, in Ref~\cite{Detmold:2015aaa}, a precise calculation of the $\Lambda_b \to \Lambda_c$ and $\Lambda_b \to p$ form factors using lattice QCD with $2+1$ dynamical flavors has been done and the SM prediction of the differential and integrated decay rates of $\Lambda_b \to \Lambda_c\,l\,\nu$ and $\Lambda_b \to p\,l\,\nu$ decays have been reported.  In this paper, we use the most general effective Lagrangian in the presence of NP and study the effect of various NP couplings on different observables such as differential decay distribution, ratio of branching ratios, forward backward asymmetries, and the convexity parameter for $\Lambda_b \to \Lambda_c\,l\,\nu$ and $\Lambda_b \to p\,l\,\nu$ decays in a model independent way. Although we adopt the same approach, our treatment differs significantly from Ref.~\cite{Shivashankara:2015cta}. We treat $b \to u$ and $b \to c$ semileptonic decays together in the same framework and perform a combined analysis using the constraints coming from $R_D$, $R_{D^{\ast}}$, and $R_{\pi}^l$ to the end in determining the possible ranges in each observables. Again, for the $\Lambda_b \to \Lambda_c$ and $\Lambda_b \to p$ transition form factors, we use the most precise lattice calculations of Ref.~\cite{Detmold:2015aaa}.

This paper is organized as follows. In section.~\ref{ehha}, we start with the most general expression for the effective Lagrangian for the $b \to (c,,u)\,l\,\nu$ transition decays in the presence of NP. A brief discussion on $\Lambda_b \to \Lambda_c$ and $\Lambda_b \to p$ transition form factors are also presented. All the relevant formulas pertinent for our numerical calculation are reported in section~\ref{ehha}. We define several observables such as differential branching ratio, ratio of branching ratios, forward backward asymmetries, and convexity parameters for the $\Lambda_b \to \Lambda_c\,\tau\nu$ and $\Lambda_b \to p\,\tau\nu$ decay modes. In section~\ref{rd}, we start with various input parameters that are used for our analysis. The SM prediction and the effect of various NP couplings on all the observables for the $\Lambda_b \to \Lambda_c\,\tau\nu$ and $\Lambda_b \to p\,\tau\nu$ decay modes are presented in section~\ref{rd}.  We present a brief summary of our results and conclude in section~\ref{con}.

\section{Effective Lagrangian and helicity amplitudes}
\label{ehha}
In the presence of NP, the effective weak Lagrangian for the $b \to q^{\prime}\,l\,\nu$ transition decays, where $q^{\prime}$ is either a $u$ quark or a $c$ quark, can be written as~\cite{Bhattacharya, Cirigliano}
\begin{eqnarray}
\mathcal L_{\rm eff} &=&
-\frac{4\,G_F}{\sqrt{2}}\,V_{q^\prime b}\,\Bigg\{(1 + V_L)\,\bar{l}_L\,\gamma_{\mu}\,\nu_L\,\bar{q^\prime}_L\,\gamma^{\mu}\,b_L +
V_R\,\bar{l}_L\,\gamma_{\mu}\,\nu_L\,\bar{q^\prime}_R\,\gamma^{\mu}\,b_R \nonumber \\
&&+
\widetilde{V}_L\,\bar{l}_R\,\gamma_{\mu}\,\nu_R\,\bar{q^\prime}_L\,\gamma^{\mu}\,b_L +
\widetilde{V}_R\,\bar{l}_R\,\gamma_{\mu}\,\nu_R\,\bar{q^\prime}_R\,\gamma^{\mu}\,b_R \nonumber \\
&&+
S_L\,\bar{l}_R\,\nu_L\,\bar{q^\prime}_R\,b_L +
S_R\,\bar{l}_R\,\nu_L\,\bar{q^\prime}_L\,b_R \nonumber \\
&&+
\widetilde{S}_L\,\bar{l}_L\,\nu_R\,\bar{q^\prime}_R\,b_L +
\widetilde{S}_R\,\bar{l}_L\,\nu_R\,\bar{q^\prime}_L\,b_R \nonumber \\
&&+ 
T_L\,\bar{l}_R\,\sigma_{\mu\nu}\,\nu_L\,\bar{q^\prime}_R\,\sigma^{\mu\nu}\,b_L +
\widetilde{T}_L\,\bar{l}_L\,\sigma_{\mu\nu}\,\nu_R\,\bar{q^\prime}_L\,\sigma^{\mu\nu}\,b_R\Bigg\} + {\rm h.c.}\,,
\end{eqnarray}
where $G_F$ is the Fermi constant, $ V_{q^\prime b}$ is the relevant CKM Matrix element, and $(q^{\prime},\,b,\,l,\,\nu)_{R,\,L} = \Big(\frac{1 \pm \gamma_5}{2}\Big)\,(q^{\prime},\,b,\,l,\,\nu)$. The NP couplings, associated with new vector, scalar, and tensor interactions, denoted by $V_{L,R}$, $S_{L,R}$, and $ T_{L}$ involve
left-handed neutrinos, whereas, the NP couplings denoted by $\widetilde{V}_{L,R}$, $\widetilde{S}_{L,R}$, and $\widetilde{T}_{L}$ involve right-handed neutrinos.
We consider NP contributions coming from vector and scalar type of interactions only. We neglect the contributions coming from NP couplings that involves right-handed neutrinos, i.e,
$\widetilde{V}_{L,R} = \widetilde{S}_{L,R}$ = $\widetilde{T}_{L} = 0$.  All the NP couplings are assumed to be real for our analysis. With these assumptions and retaining the same notation as in Ref.~\cite{Dutta:2013qaa}, we obtain
\begin{eqnarray}
\label{leff}
\mathcal L_{\rm eff} &=&
-\frac{G_F}{\sqrt{2}}\,V_{q^\prime b}\,\Bigg\{G_V\,\bar{l}\,\gamma_{\mu}\,(1 - \gamma_5)\,\nu_l\,\bar{q^\prime}\,\gamma^{\mu}\,b -
G_A\,\bar{l}\,\gamma_{\mu}\,(1 - \gamma_5)\,\nu_l\,\bar{q^\prime}\,\gamma^{\mu}\,\gamma_5\,b \nonumber \\
&&+
G_S\,\bar{l}\,(1 - \gamma_5)\,\nu_l\,\bar{q^\prime}\,b - G_P\,\bar{l}\,(1 - \gamma_5)\,\nu_l\,\bar{q^\prime}\,\gamma_5\,b \Bigg\} + {\rm h.c.}\,,
\end{eqnarray}
where 
\begin{eqnarray*} 
&&G_V = 1 + V_L + V_R\,,\qquad\qquad
G_A = 1 + V_L - V_R\,, \qquad\qquad
G_S = S_L + S_R\,,\qquad\qquad
G_P = S_L - S_R\,.
\end{eqnarray*}
The SM contribution can be obtained once we set $V_{L,R} = S_{L,R} = 0$ in Eq.~(\ref{leff}).

In order to compute the branching fractions and other observables for $\Lambda_b \to \Lambda_c\,l\,\nu$ and $\Lambda_b \to p\,l\,\nu$ decay modes, we need to find various hadronic form factors that parametrizes the hadronic matrix elements of vector~(axial vector) and scalar~(pseudoscalar) currents between the two spin half baryons. 
The hadronic matrix elements of vector and axial vector currents between two spin half baryons $B_1$ and $B_2$ can be parametrized in terms of various form factors as
\begin{eqnarray}
&&M_{\mu}^V = \langle B_2,\lambda_2 | J_{\mu}^V | B_1,\lambda_1\rangle = \bar{u}_2(p_2,\lambda_2)\Big[f_1(q^2)\gamma_{\mu} + if_2(q^2)\sigma_{\mu\nu}\,q^{\nu} + f_3(q^2)q_{\mu}\Big]u_1(p_1,\lambda_1)\,,\nonumber \\
&&M_{\mu}^A= \langle B_2,\lambda_2 | J_{\mu}^A | B_1,\lambda_1\rangle = \bar{u}_2(p_2,\lambda_2)\Big[g_1(q^2)\gamma_{\mu} + ig_2(q^2)\sigma_{\mu\nu}\,q^{\nu} + g_3(q^2)q_{\mu}\Big]\,\gamma_5\,u_1(p_1,\lambda_1)\,,
\end{eqnarray}
where $q^{\mu} = (p_1 - p_2)^{\mu}$ is the four momentum transfer, $\lambda_1$ and $\lambda_2$ are the helicities of the parent and daughter baryons, respectively and $\sigma_{\mu\nu} = \frac{i}{2}\,[\gamma_{\mu}, \gamma_{\nu}]$. Here $B_1 = \Lambda_b$ and $B_2 = (\Lambda_c,\,p)$, respectively. In order to find the hadronic matrix elements of scalar and pseudoscalar currents, we use the equation of motion. That is
\begin{eqnarray}
&&\langle B_2,\lambda_2 | \bar{q}^{\prime}\,b | B_1,\lambda_1\rangle = \bar{u}_2(p_2,\lambda_2)\Big[f_1(q^2)\,\frac{\not q}{m_b - m_{q^{\prime}}} +  f_3(q^2)\,\frac{q^2}{m_b - m_{q^{\prime}}} \Big]u_1(p_1,\lambda_1)\,,\nonumber \\
&&\langle B_2,\lambda_2 |  \bar{q}^{\prime}\,\gamma_5\,b  | B_1,\lambda_1\rangle = \bar{u}_2(p_2,\lambda_2)\Big[-g_1(q^2)\frac{\not q}{m_b + m_{q^{\prime}}} -  g_3(q^2)\,\frac{q^2}{m_b + m_{q^{\prime}}} \Big]\,\gamma_5\,u_1(p_1,\lambda_1)\,,
\end{eqnarray}
where $m_b$ is the mass of $b$ quark and $m_{q^{\prime}}$ is the mass of $q^{\prime} = (u,\,c)$ quarks evaluated at renormalization scale $\mu = m_b$, respectively.
The various form factors $f_i$'s and $g_i$'s are related to the helicity form factors $f_{+,\perp,0}$ and $g_{+,\perp,0}$ as follows~\cite{Detmold:2015aaa}:
\begin{eqnarray}
&&f_+(q^2) = f_1(q^2) - \frac{q^2}{m_{B_1} + m_{B_2}}\,f_2(q^2)\,, \nonumber \\
&&f_{\perp}(q^2) = f_1(q^2) - (m_{B_1} + m_{B_2})\,f_2(q^2)\,, \nonumber \\
&&f_0(q^2) =  f_1(q^2) + \frac{q^2}{m_{B_1} - m_{B_2}}\,f_3(q^2)\,, \nonumber \\
&&g_+(q^2) = g_1(q^2) +\frac{q^2}{m_{B_1} - m_{B_2}}\,g_2(q^2)\,, \nonumber \\
&&g_{\perp}(q^2) = g_1(q^2) + (m_{B_1} - m_{B_2})\,g_2(q^2)\,, \nonumber \\
&&g_0(q^2) =  g_1(q^2) + \frac{q^2}{m_{B_1} + m_{B_2}}\,g_3(q^2)\,,
\end{eqnarray}
where $m_{B_1}$ and $m_{B_2}$ are the masses of $B_1$ and $B_2$ baryons, respectively.
For the various helicity form factors we have used the formula given in Ref~.\cite{Detmold:2015aaa}. The relevant equations pertinent for our calculation are as follows:
\begin{eqnarray}
&&f(q^2) = \frac{1}{1-q^2/(m_{\rm pole}^f)^2}\,\Big[a_0^f + a_1^f\,z(q^2)\Big]\,,
\end{eqnarray}
where $m_{\rm pole}^f$ is pole mass. Here $f$ represents $f_{+,{\perp},\,0}$ and $g_{+,{\perp},\,0}$, respectively. The numerical values of $m_{\rm pole}^f$, $a_0^f$, and $a_1^f$ relevant for our calculation are taken from Ref.~\cite{Detmold:2015aaa}. The expansion parameter $z$ is defined as
\begin{eqnarray}
&&z(q^2) = \frac{\sqrt{t_+ - q^2} - \sqrt{t_+ - t_0}}{\sqrt{t_+ - q^2} + \sqrt{t_+ - t_0}}\,,
\end{eqnarray}
where $t_+ = (m_{B_1} + m_{B_2})^2$ and $t_0 = (m_{B_1} - m_{B_2})^2$, respectively. For more details, we refer to Ref.~\cite{Detmold:2015aaa}. 
We now proceed to discuss the helicity amplitudes for these baryonic $b \to (c,\,u)\,l\,\nu$ decay modes. The helicity amplitudes can be defined by~\cite{Korner,Gutsche:2015mxa}
\begin{eqnarray}
&&H_{\lambda_2\,\lambda_W}^{V/A} = M_{\mu}^{V/A}(\lambda_2)\,\epsilon^{\dagger^{\mu}}(\lambda_W)\,,
\end{eqnarray}
where $\lambda_2$ and $\lambda_W$ denote the helicities of the daughter baryon and $W^-_{\rm off - shell}$, respectively. The total left - chiral helicity amplitude can be written as
\begin{eqnarray}
&&H_{\lambda_2\,\lambda_W} = H_{\lambda_2\,\lambda_W}^{V} - H_{\lambda_2\,\lambda_W}^{A}\,. 
\end{eqnarray}
In terms of the various form factors and the NP couplings, the helicity amplitudes can be written as~\cite{Shivashankara:2015cta}
\begin{eqnarray}
&&H_{\frac{1}{2}\,0}^V = G_V\,\frac{\sqrt{Q_-}}{\sqrt{q^2}}\,\Big[(m_{B_1} + m_{B_2})\,f_1(q^2) - q^2\,f_2(q^2)\Big]\,, \nonumber \\
&&H_{\frac{1}{2}\,0}^A = G_A\,\frac{\sqrt{Q_+}}{\sqrt{q^2}}\,\Big[(m_{B_1} - m_{B_2})\,g_1(q^2) + q^2\,g_2(q^2)\Big]\,, \nonumber \\
&&H_{\frac{1}{2}\,1}^V = G_V\,\sqrt{2\,Q_-}\,\Big[-f_1(q^2) + (m_{B_1} + m_{B_2})\,f_2(q^2)\Big]\,, \nonumber \\
&&H_{\frac{1}{2}\,1}^A = G_A\,\sqrt{2\,Q_+}\,\Big[-g_1(q^2) - (m_{B_1} - m_{B_2})\,g_2(q^2)\Big]\,, \nonumber \\
&&H_{\frac{1}{2}\,t}^V = G_V\,\frac{\sqrt{Q_+}}{\sqrt{q^2}}\,\Big[(m_{B_1} - m_{B_2})\,f_1(q^2) + q^2\,f_3(q^2)\Big]\,, \nonumber \\
&&H_{\frac{1}{2}\,t}^A = G_A\,\frac{\sqrt{Q_-}}{\sqrt{q^2}}\,\Big[(m_{B_1} + m_{B_2})\,g_1(q^2) - q^2\,g_3(q^2)\Big]\,, 
\end{eqnarray}
where $Q_{\pm} = (m_{B_1} \pm m_{B_2})^2 - q^2$. Either from parity or from explicit calculation, one can show that $H^V_{-\lambda_2\, -\lambda_W} = H^V_{\lambda_2\, \lambda_W}$ and $H^A_{-\lambda_2\, -\lambda_W} = - H^A_{\lambda_2\, \lambda_W}$. Similarly, the scalar and pseudoscalar helicity amplitudes associated with the NP couplings $G_S$ and $G_P$ can be written as~\cite{Shivashankara:2015cta}
\begin{eqnarray}
&&H_{\frac{1}{2}\,0}^{SP} = H_{\frac{1}{2}\,0}^{S} - H_{\frac{1}{2}\,0}^{P}\,, \nonumber \\
&&H_{\frac{1}{2}\,0}^{S} = G_S\,\frac{\sqrt{Q_+}}{m_b - m_{q^{\prime}}}\,\Big[(m_{B_1} - m_{B_2})\,f_1(q^2) + q^2\,f_3(q^2)\Big]\,, \nonumber \\
&&H_{\frac{1}{2}\,0}^{P} = G_P\,\frac{\sqrt{Q_-}}{m_b + m_{q^{\prime}}}\,\Big[(m_{B_1} + m_{B_2})\,g_1(q^2) - q^2\,g_3(q^2)\Big]\,.
\end{eqnarray}
Moreover,  we have $H_{\lambda_2\,\lambda_{\rm NP}}^S = H_{-\lambda_2\,-\lambda_{\rm NP}}^S$ and $H_{\lambda_2\,\lambda_{\rm NP}}^P = -H_{-\lambda_2\,-\lambda_{\rm NP}}^P$ from parity argument or from explicit calculation.

We follow Ref.~\cite{Shivashankara:2015cta} and write the differential angular distribution for the three body $B_1 \to B_2\,l\,\nu$ decays in the presence of NP as
\begin{eqnarray}
\label{dtdq2dcth}
&&\frac{d\Gamma(B_1 \to B_2\,l\,\nu)}{dq^2\,d\cos\theta_l}= N\,\Big(1-\frac{m_l^2}{q^2}\Big)^2\Big[A_1 + \frac{m_l^2}{q^2}\,A_2 + 2\,A_3 + \frac{4\,m_l}{\sqrt{q^2}}\,A_4\Big]\,,
\end{eqnarray}
where
\begin{eqnarray}
N &=& \frac{G_F^2\,|V_{q^{\prime}\,b}|^2\,q^2\,|\vec{p}_{B_2}|}{512\,\pi^3\,m_{B_1}^2}\,,
\nonumber \\
A_1 &=& 2\,\sin^2\theta_l\,\Big(H_{\frac{1}{2}\,0}^2 + H_{-\frac{1}{2}\,0}^2\Big) +\Big(1 - \cos\theta_l\Big)^2\,H_{\frac{1}{2}\,1}^2 + \Big(1 + \cos\theta_l\Big)^2\,H_{-\frac{1}{2}\,-1}^2\,, \nonumber \\
A_2 &=& 2\,\cos^2\theta_l\,\Big(H_{\frac{1}{2}\,0}^2 + H_{-\frac{1}{2}\,0}^2\Big) + \sin^2\theta_l\,\Big(H_{\frac{1}{2}\,1}^2 + H_{-\frac{1}{2}\,-1}^2\Big) + 2\,\Big(H_{\frac{1}{2}\,t}^2 + H_{-\frac{1}{2}\,t}^2\Big) - \nonumber \\
&&4\,\cos\theta_l\,\Big(H_{\frac{1}{2}\,t}\,H_{\frac{1}{2}\,0} + H_{-\frac{1}{2}\,t}\,H_{-\frac{1}{2}\,0}\Big)\,, \nonumber \\
A_3 &=& (H^{SP}_{\frac{1}{2}\,0})^2 + (H^{SP}_{-\frac{1}{2}\,0})^2\,, \nonumber \\
A_4 &=& -\cos\theta_l\,\Big(H_{\frac{1}{2}\,0}\,H^{SP}_{\frac{1}{2}\,0} + H_{-\frac{1}{2}\,0}\,H^{SP}_{-\frac{1}{2}\,0}\Big) + \Big(H_{\frac{1}{2}\,t}\,H^{SP}_{\frac{1}{2}\,0} + H_{-\frac{1}{2}\,t}\,H^{SP}_{-\frac{1}{2}\,0}\Big)\,.
\end{eqnarray}
Here $ |\vec{p}_{B_2}| = \sqrt{\lambda(m_{B_1}^2,\,m_{B_2}^2,\,q^2)}/2\,m_{B_1}$ is the momentum of the outgoing baryon $B_2$, where $\lambda(a,\,b,\,c) = a^2 + b^2 + c^2 - 2\,(a\,b + b\,c + c\,a)$. We denote $\theta_l$ as the angle between the daughter baryon $B_2$ and the lepton three momentum vector in the $q^2$ rest frame.
The differential decay rate can be obtained by integrating out $\cos\theta_l$ from Eq.~(\ref{dtdq2dcth}), i.e,
\begin{eqnarray}
&&\frac{d\Gamma(B_1 \to B_2\,l\,\nu)}{dq^2}= \frac{8\,N}{3}\,\Big(1-\frac{m_l^2}{q^2}\Big)^2\Big[B_1 + \frac{m_l^2}{2\,q^2}\,B_2 + \frac{3}{2}\,B_3 + \frac{3\,m_l}{\sqrt{q^2}}\,B_4\Big]\,,
\end{eqnarray}
where
\begin{eqnarray}
&&B_1 = H_{\frac{1}{2}\,0}^2 + H_{-\frac{1}{2}\,0}^2 + H_{\frac{1}{2}\,1}^2 + H_{-\frac{1}{2}\,-1}^2\,, \nonumber \\
&&B_2 = H_{\frac{1}{2}\,0}^2 + H_{-\frac{1}{2}\,0}^2 + H_{\frac{1}{2}\,1}^2 + H_{-\frac{1}{2}\,-1}^2 + 3\,\Big(H_{\frac{1}{2}\,t}^2 + H_{-\frac{1}{2}\,t}^2\Big)\,, \nonumber \\
&&B_3 = (H^{SP}_{\frac{1}{2}\,0})^2 + (H^{SP}_{-\frac{1}{2}\,0})^2\,, \nonumber \\
&&B_4 = H_{\frac{1}{2}\,t}\,H^{SP}_{\frac{1}{2}\,0} + H_{-\frac{1}{2}\,t}\,H^{SP}_{-\frac{1}{2}\,0}\,.
\end{eqnarray}
We define several observables such as ratio of branching ratios $R_{\Lambda_c}$, $R_p$, and ratio of partially integrated branching ratios $R_{\Lambda_c\,p}^{\mu}$ for the two decay modes such that
\begin{eqnarray}
\label{rdds}
&& R_{\Lambda_c}=\frac{\mathcal B(\Lambda_b \to \Lambda_c\tau^- \bar{\nu}_\tau)}{\mathcal B( \Lambda_b \to \Lambda_c\, l^- \bar{\nu}_l)} \,, \nonumber\\
&& R_{p}=\frac{\mathcal B(\Lambda_b \to p\tau^- \bar{\nu}_\tau)}{\mathcal B(\Lambda_b \to p\, l^- \bar{\nu}_l)}\,, \nonumber \\
&& R_{\Lambda_c\,p}^{\mu} = \frac{\int_{15\,{\rm GeV^2}}^{q^2_{\rm max}}\frac{d\Gamma(\Lambda_b \to p\,\mu\,\nu)}{dq^2}\,dq^2}{\int_{7\,{\rm GeV^2}}^{q^2_{\rm max}}\frac{d\Gamma(\Lambda_b \to \Lambda_c\,\mu\,\nu)}{dq^2}\,dq^2}\,.
\end{eqnarray}
We have also defined several $q^2$ dependent observables such as differential branching fractions ${\rm DBR}(q^2)$, ratio of branching fractions $R(q^2)$, forward backward asymmetries $A_{\rm FB}(q^2)$, and the convexity parameter $C_F^l(q^2)$ for these two baryonic decay modes. Those are
\begin{eqnarray}
&&{\rm DBR}(q^2) = \Big(\frac{d\Gamma}{dq^2}\Big)\Big/\Gamma_{\rm tot}\,, \qquad\qquad
R(q^2)=\frac{DBR(q^2)\Big(B_1 \to B_2\,\tau\,\nu\Big)}{DBR(q^2)\Big(B_1 \to B_2\,l\,\nu\Big)}\,, 
\nonumber\\
&&A_{\rm FB}(q^2)=\Bigg\{\Big(\int_{-1}^{0}-\int_{0}^{1}\Big)d\cos \theta_l\frac{d\Gamma}{dq^2\,d\cos\theta_l}\Bigg\}\Big/\frac{d\Gamma}{dq^2}\,, \qquad\qquad
C_F^l(q^2) = \frac{1}{\mathcal H_{\rm tot}}\,\frac{d^2\,W(\theta)}{d(\cos\theta)^2}\,,
\end{eqnarray}
where
\begin{eqnarray}
&&W(\theta) = \frac{3}{8}\,\Big[A_1 + \frac{m_l^2}{q^2}\,A_2 + 2\,A_3 + \frac{4\,m_l}{\sqrt{q^2}}\,A_4\Big]\,, \nonumber \\
&& \mathcal H_{\rm tot} = \int\,d(\cos\theta)\,W(\theta)\,, \nonumber \\
&&\frac{d^2\,W(\theta)}{d(\cos\theta)^2} = \frac{3}{4}\,\Big(1 - \frac{m_l^2}{q^2}\Big)\,\Big[H_{\frac{1}{2}\,1}^2 + H_{-\frac{1}{2}\,-1}^2 - 2\,\Big(H_{\frac{1}{2}\,0}^2 + H_{-\frac{1}{2}\,0}^2\Big)\Big]\,.
\end{eqnarray}
We want to mention that the observable $\frac{d^2\,W(\theta)}{d(\cos\theta)^2}$ is independent of the new scalar couplings $S_L$ and $S_R$. It only depends on the new vector couplings $V_L$ and $V_R$. Hence, once NP is established, this observable can be used to distinguish between the vector and scalar type of NP interactions. We now proceed to discuss the results of our analysis.

\section{Results and Discussion}
\label{rd}
For definiteness, we first present all the inputs that are pertinent for our calculation. For the quark, lepton, and the baryon masses, we use $m_b(m_b) = 4.18\,{\rm GeV}$, $m_c(m_b) = 0.91\,{\rm GeV}$, $m_e = 0.510998928 \times 10^{-3}\,{\rm GeV}$, $m_{\mu} = 0.1056583715\,{\rm GeV}$, $m_{\tau} = 1.77682\,{\rm GeV}$, $m_p = 0.938272046\,{\rm GeV}$, $m_{\Lambda_b} = 5.61951\,{\rm GeV}$, $m_{\Lambda_c} = 2.28646\,{\rm GeV}$~\cite{pdg}. For the mean life time of $\Lambda_b$ baryon, we use $\tau_{\Lambda_b} = (1.466 \pm 0.010)\times 10^{-12}\,{\rm s}$~\cite{pdg}. For the CKM matrix element $|V_{cb}|$, we have used the value $|V_{cb}| = (39.5 \pm 0.8) \times 10^{-3}$~\cite{pdg}. Very recently, LHCb measured the partially integrated decay rates of $\Lambda_b^0$ baryon to decay into the $p\,\mu\,\nu$ final state relative to the $\Lambda_c^+\,\mu\,\nu$ final state to be
\begin{eqnarray}
R_{\Lambda_c\,p}^{\mu} = \frac{\int_{15\,{\rm GeV^2}}^{q^2_{\rm max}}\frac{d\Gamma(\Lambda_b \to p\,\mu\,\nu)}{dq^2}\,dq^2}{\int_{7\,{\rm GeV^2}}^{q^2_{\rm max}}\frac{d\Gamma(\Lambda_b \to \Lambda_c\,\mu\,\nu)}{dq^2}\,dq^2} = (1.00 \pm 0.04 \pm 0.08) \times 10^{-2}
\end{eqnarray}
 and put  constraint on the ratio $|V_{ub}|/|V_{cb}| = 0.083 \pm 0.004 \pm 0.004$~\cite{Aaij:2015bfa}. A value of $|V_{ub}| = (32.7 \pm 2.3) \times 10^{-4}$~\cite{Aaij:2015bfa,Fiore:2015cmx} is obtained using the theoretical calculations and the extracted value of $|V_{cb}|$ from exclusive $B$ decays. This measurement of $|V_{ub}|$ using baryonic decay channel is in very good agreement with the exclusively measured world average from Ref.~\cite{Amhis:2014hma}. However, it disagrees with the inclusive measurement~\cite{pdg} at a significance level of $3.5\sigma$. A very precise calculation of the $\Lambda_b \to \Lambda_c$ and $\Lambda_b \to p$ hadronic form factors using lattice QCD with $2 + 1$ dynamical flavors relevant for the determination of CKM elements $|V_{cb}|$ and $|V_{ub}|$ is very recently done in Ref.~\cite{Detmold:2015aaa}. The relevant parameters for the form factor calculation are given in Table.~\ref{tab1} and Table.~\ref{tab2}. We also report the most important experimental input parameters $R_D$, $R_{D^{\ast}}$, and $R_{\pi}^l$ with their uncertainties in Table.~\ref{tab4}. The errors in Eq.~(\ref{rdrds}) are added in quadrature. Let us now proceed to discuss the results that are obtained within the SM. 
\begin{table}
\begin{center}
\begin{tabular}{|c|c|c||c|c|c|}
\hline
$f$&$m^f_{\rm pole}(\Lambda_b \to \Lambda_c)$&$m^f_{\rm pole}(\Lambda_b \to p)$&$f$&$m^f_{\rm pole}(\Lambda_b \to \Lambda_c)$&$m^f_{\rm pole}(\Lambda_b \to p)$ \\[0.2cm]
\hline
$ f_+,\,f_{\perp}$ & $6.332$ &$5.325$& $  g_+,\,g_{\perp} $ & $6.768$&$5.706$ \\[0.2cm]
$ f_0 $ & $6.725$&$5.655$&$ g_0 $ & $6.276$ &$5.279$ \\[0.2cm]
\hline
\end{tabular}
\end{center}
\caption{Masses~(in GeV) of the relevant form factor pole taken from Ref.~\cite{Detmold:2015aaa}}
\label{tab1}
\end{table}
\begin{table}
\begin{center}
\begin{tabular}{|c|c|c||c|c|c|}
\hline
Parameter&$\Lambda_b \to p$  &$ \Lambda_b \to \Lambda_c$ &Parameter&$\Lambda_b \to p$  &$ \Lambda_b \to \Lambda_c$ \\[0.2cm]
\hline
$ a_0^{f_+}$ & $0.43\pm 0.03$ &$0.8137\pm 0.0181$&$ a_0^{g_+}$ & $0.3718\pm 0.0194$ &$0.6876\pm 0.0084$ \\[0.1cm]
$ a_1^{f_+} $ & $-1.4578\pm 0.4178$&$-8.5673\pm 0.8444$&$ a_1^{g_+} $ & $-1.4561\pm 0.3280$&$-6.5556\pm 0.4713$ \\[0.1cm]
$ a_0^{f_0}  $ & $0.3981\pm 0.0245$&$0.7494\pm 0.0132$ &$ a_0^{g_0}  $ & $0.4409\pm 0.0278$&$0.7446\pm 0.0156$ \\[0.1cm]
$  a_1^{f_0} $ & $-1.3575\pm 0.3869$ &$-7.2530\pm 0.8114$&$  a_1^{g_0} $ & $-1.7273\pm 0.3684$ &$-7.7216\pm 0.5437$ \\[0.1cm]
$ a_0^{f_{\perp}}  $ & $0.5228\pm 0.0433$&$1.0809\pm 0.0262$ & $ a_0^{g_{\perp}}  $ & $0.3718\pm 0.0194$&$0.6876\pm 0.0084$ \\[0.1cm]
$  a_1^{f_{\perp}} $ & $-1.6943\pm 0.6834$ &$-11.6259\pm 1.5343$ &$  a_1^{g_{\perp}} $ & $-1.6839\pm 0.3882$ &$-6.7870\pm 0.5013$ \\[0.1cm]
\hline
\end{tabular}
\end{center}
\caption{Nominal form factor parameters taken from Ref.~\cite{Detmold:2015aaa}}
\label{tab2}
\end{table}
\begin{table}[htdp]
\begin{center}
\begin{tabular}{|c|c|}
\hline
\multicolumn{2}{|c|}{Ratio of branching ratios:} \\[0.2cm]
\hline
$ R_{\pi}^l $ & $0.73 \pm 0.15$ ~\cite{Fajfer:2012jt} \\[0.2cm]
$ R_D $ & $0.391 \pm 0.050 $ ~\cite{Amhis:2014hma} \\[0.2cm]
$ R_{D^{\ast}} $ & $0.322 \pm 0.022 $ ~\cite{Amhis:2014hma} \\[0.2cm]
\hline
\end{tabular}
\end{center}
\caption{Experimental input parameters}
\label{tab4}
\end{table}

The SM branching fractions and the ratio of branching fractions for the $\Lambda_b \to \Lambda_c\,l\,\nu$ and $\Lambda_b \to p\,l\,\nu$ decays are presented in Table~\ref{tab3}. There are two main sources of uncertainties. It may arise either  from not so well known input parameters such as CKM matrix elements or from hadronic input parameters such as form factors and decay constants. In order to gauge the effect of these above mentioned uncertainties on various observables, we use a random number generator and perform a random scan over all the theoretical input parameters such as CKM matrix elements, form factors, and decay constants within $1\sigma$ of their central values.
The central values reported in Table~\ref{tab3} are obtained using the central values of all the input parameters whereas, to find the $1\sigma$ range of all the parameters, we vary all the input parameters such as CKM matrix elements, the hadronic form factors, and the decay constants within $1\sigma$ from their central values. We, however, do not include the uncertainties coming from the quark mass, lepton mass, baryon mass, and the mean life time as these are not important for our analysis. Our central value for the parameter $R_{\Lambda_c}$ is exactly same as the value reported in Ref.~\cite{Detmold:2015aaa}, however, it differs slightly from the values reported in Refs.~\cite{Woloshyn:2014hka, Shivashankara:2015cta, Gutsche:2015mxa}. It is expected because we have used the lattice calculations of the form factors from Ref.~\cite{Detmold:2015aaa}. We, however, use only the nominal form factor parameters and their uncertainties in our analysis.
\begin{table}
\begin{center}
\begin{tabular}{|c|c|c|}
\hline
Observables  & Central value & $1\sigma$ range\\
\hline
$\mathcal B(\Lambda_b \to p\,l\,\nu)$ & $3.89\times 10^{-4} $ & $(1.739 - 12.870)\times 10^{-4}$  \\[0.2cm]
$\mathcal B(\Lambda_b \to p\,\tau\,\nu)$ & $2.75\times 10^{-4}$ & $(1.403 - 8.237)\times 10^{-4}$ \\[0.2cm]
$\mathcal B(\Lambda_b \to \Lambda_c\,l\,\nu)$ & $4.83\times 10^{-2}$ & $(4.316 - 5.418) \times 10^{-2}$ \\[0.2cm]
$\mathcal B(\Lambda_b \to \Lambda_c\,\tau\,\nu)$ & $1.63\times 10^{-2}$ & $(1.504 - 1.769)\times 10^{-2}$ \\[0.2cm]
$R_{\Lambda_c}$  & $0.3379$ & $(0.3203 - 0.3559)$ \\[0.2cm]
$R_p$  & $0.7071$ & $(0.588 - 0.878)$ \\[0.2cm]
$R_{\Lambda_c\,p}^{\mu}$  & $0.0101$ & $(0.0043 - 0.0333)$ \\[0.2cm]
\hline
\end{tabular}
\end{center}
\caption{Branching ratio and ratio of branching ratios within the SM.}
\label{tab3}
\end{table}

\begin{figure}[htbp]
\begin{center}
\includegraphics[width=8cm,height=5cm]{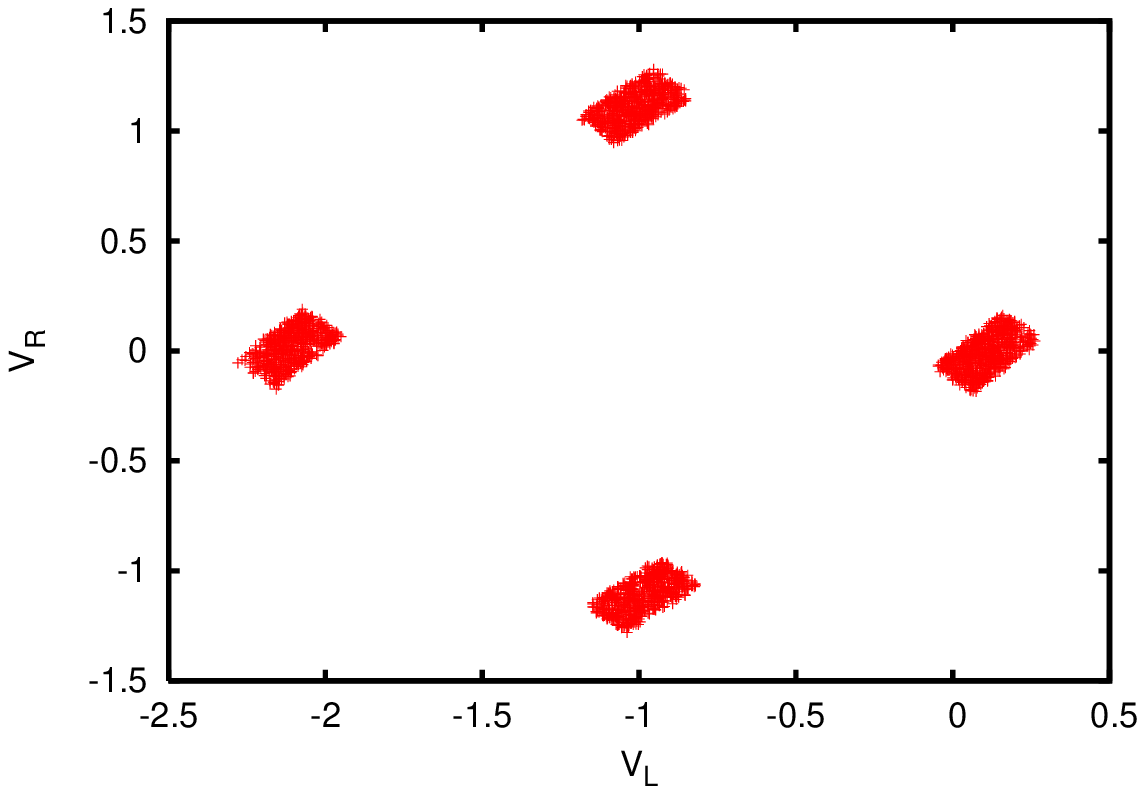}
\includegraphics[width=8cm,height=5cm]{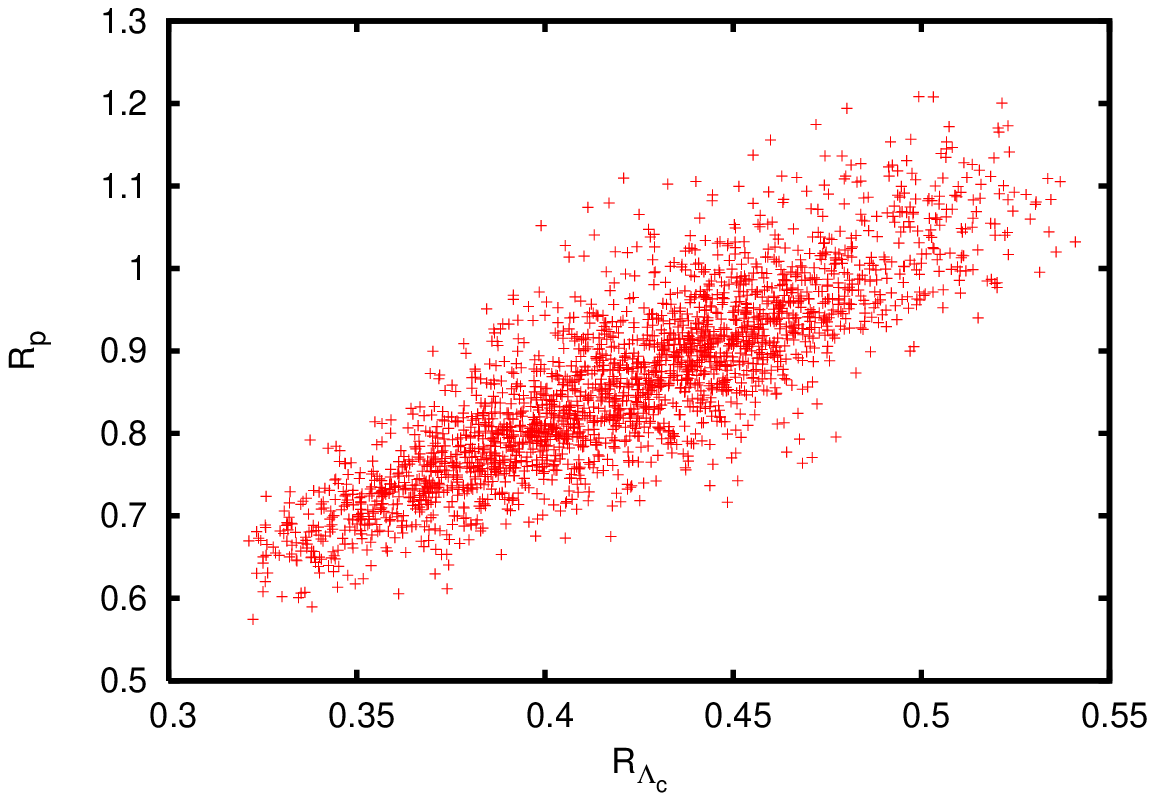}
\end{center}
\caption{Allowed regions of $V_L$ and $V_R$ obtained using the $3\sigma$ constraint coming from $R_D$, $R_{D^{\ast}}$, and $R_{\pi}^l$ are shown in the left panel and the corresponding ranges in $R_p$ and $R_{\Lambda_c}$ in
the presence of these NP couplings are shown in the right panel.}
\label{vlvr}
\end{figure}

\begin{figure}[htbp]
\begin{center}
\includegraphics[width=4cm,height=4cm]{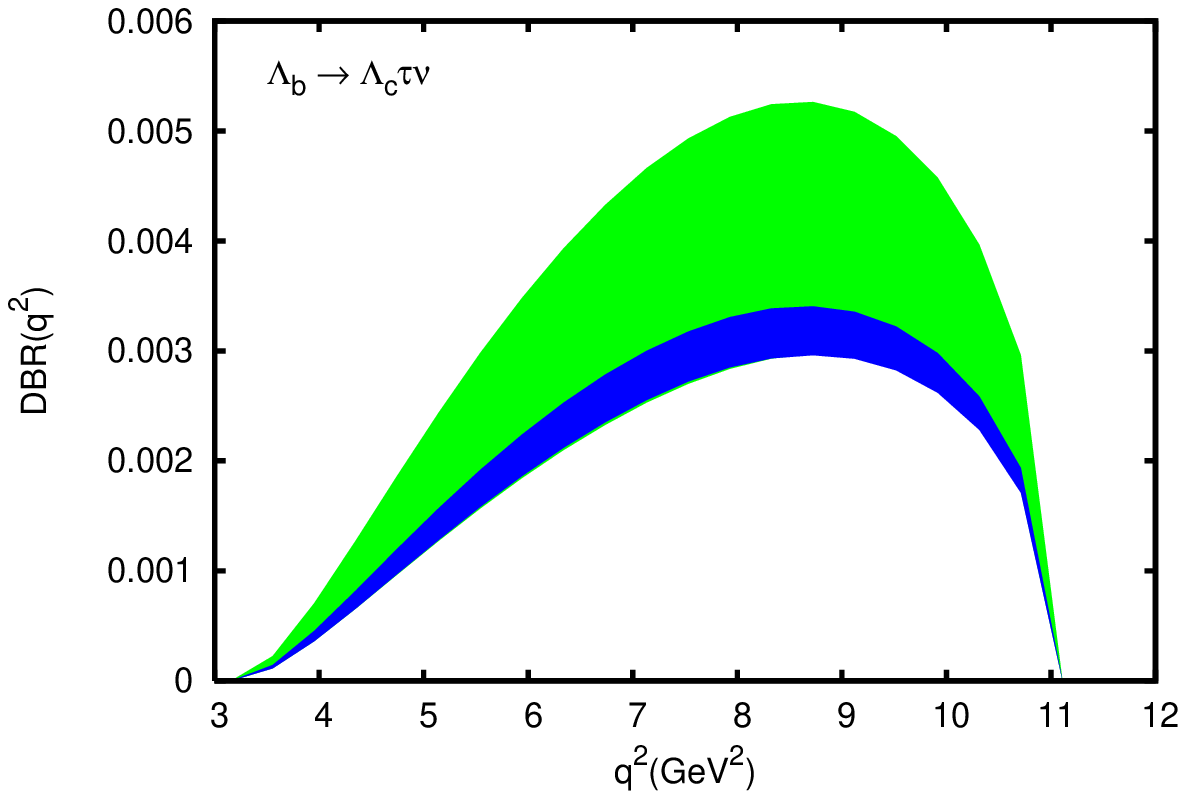}
\includegraphics[width=4cm,height=4cm]{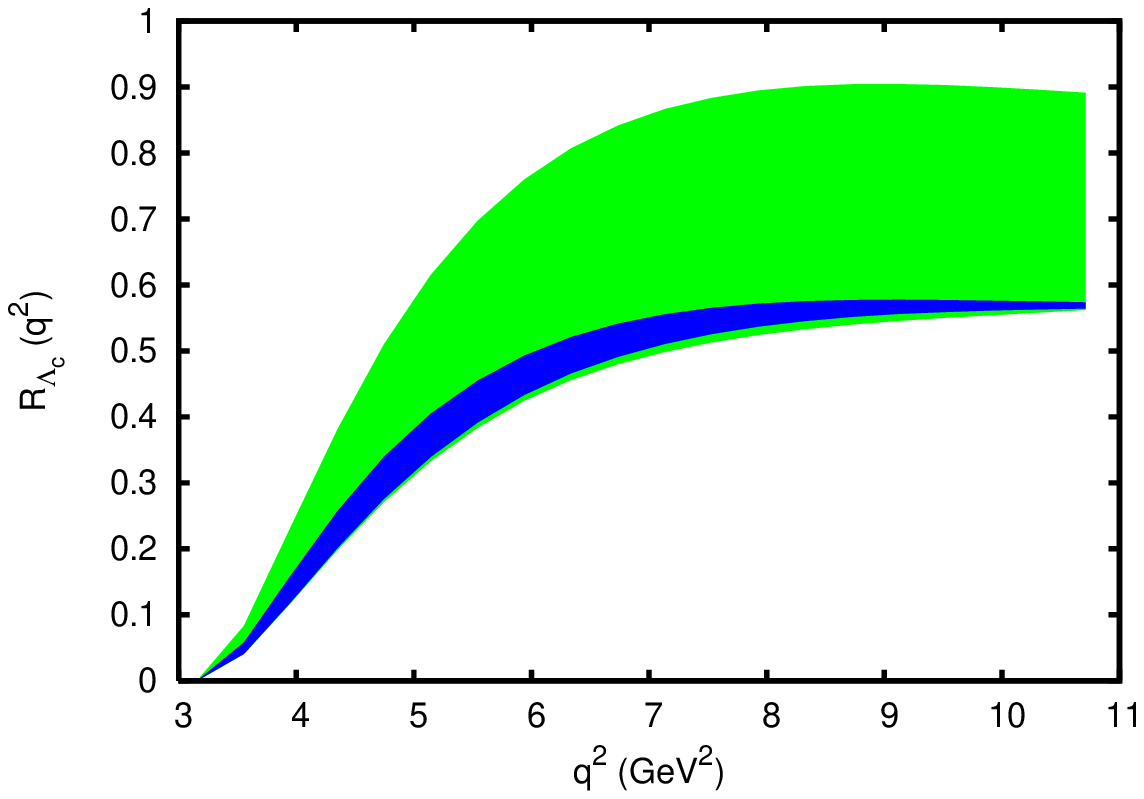}
\includegraphics[width=4cm,height=4cm]{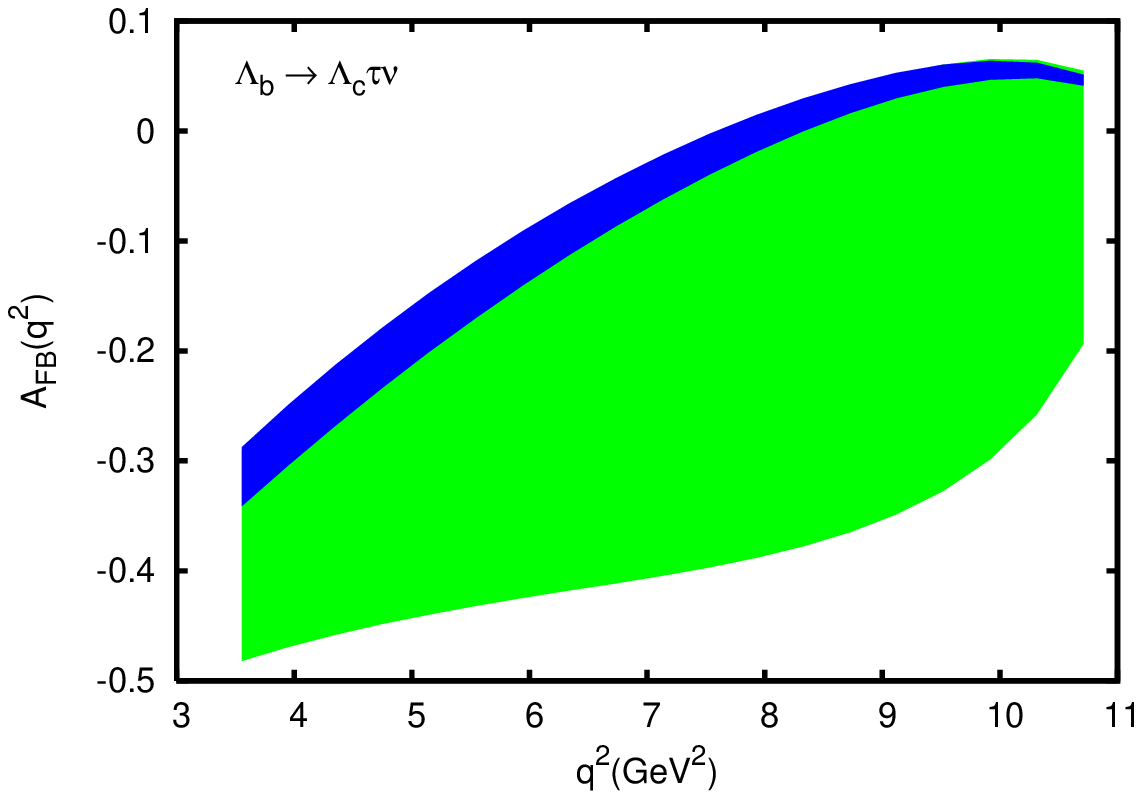}
\includegraphics[width=4cm,height=4cm]{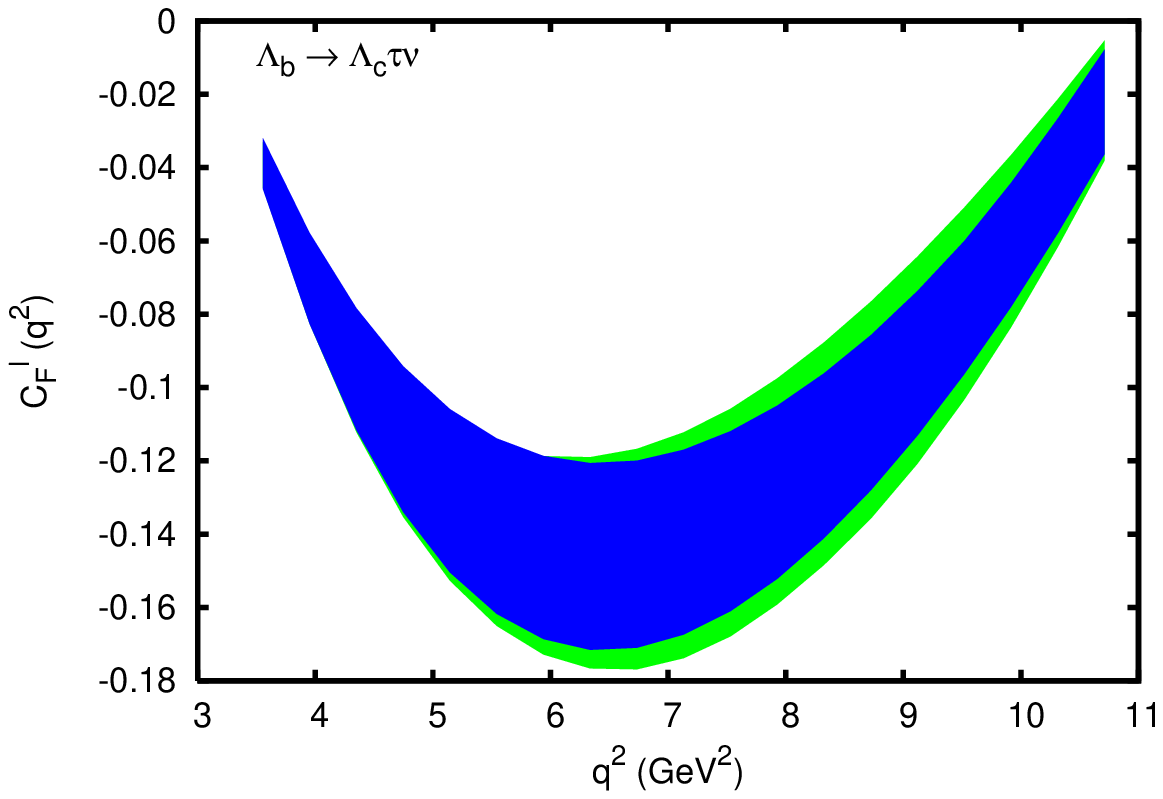}
\includegraphics[width=4cm,height=4cm]{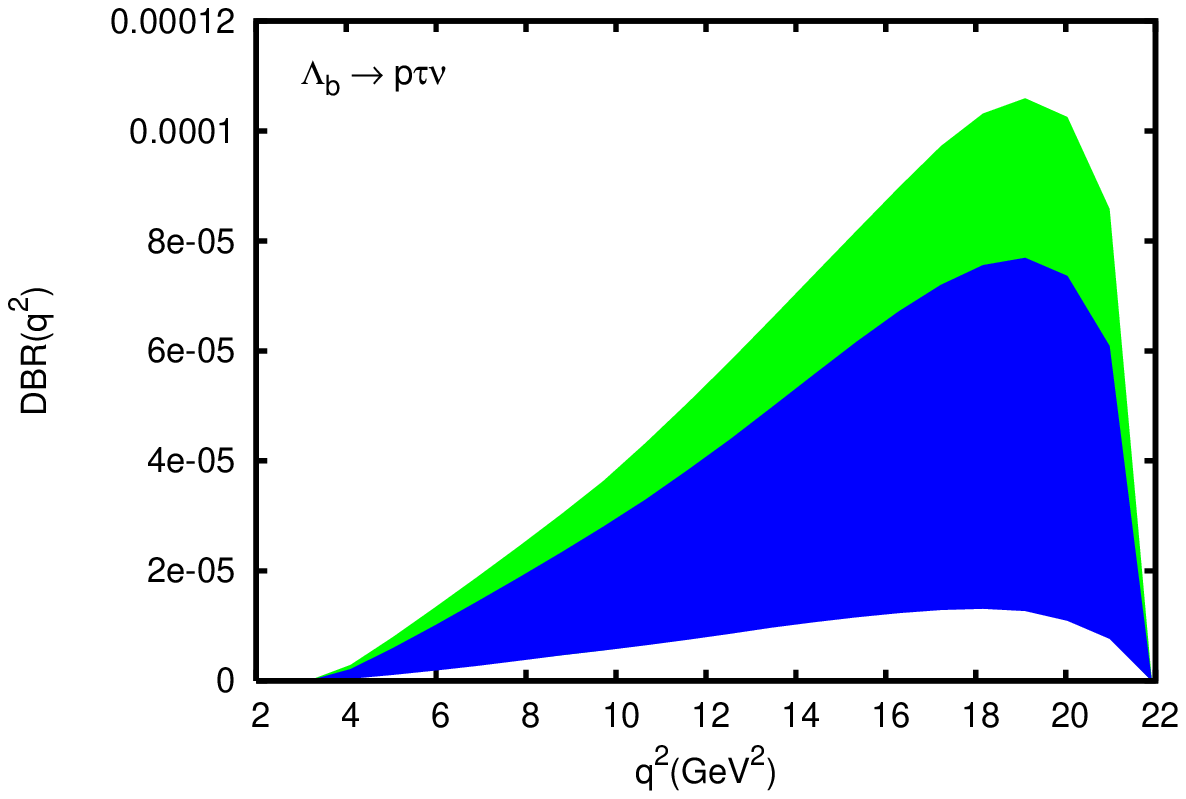}
\includegraphics[width=4cm,height=4cm]{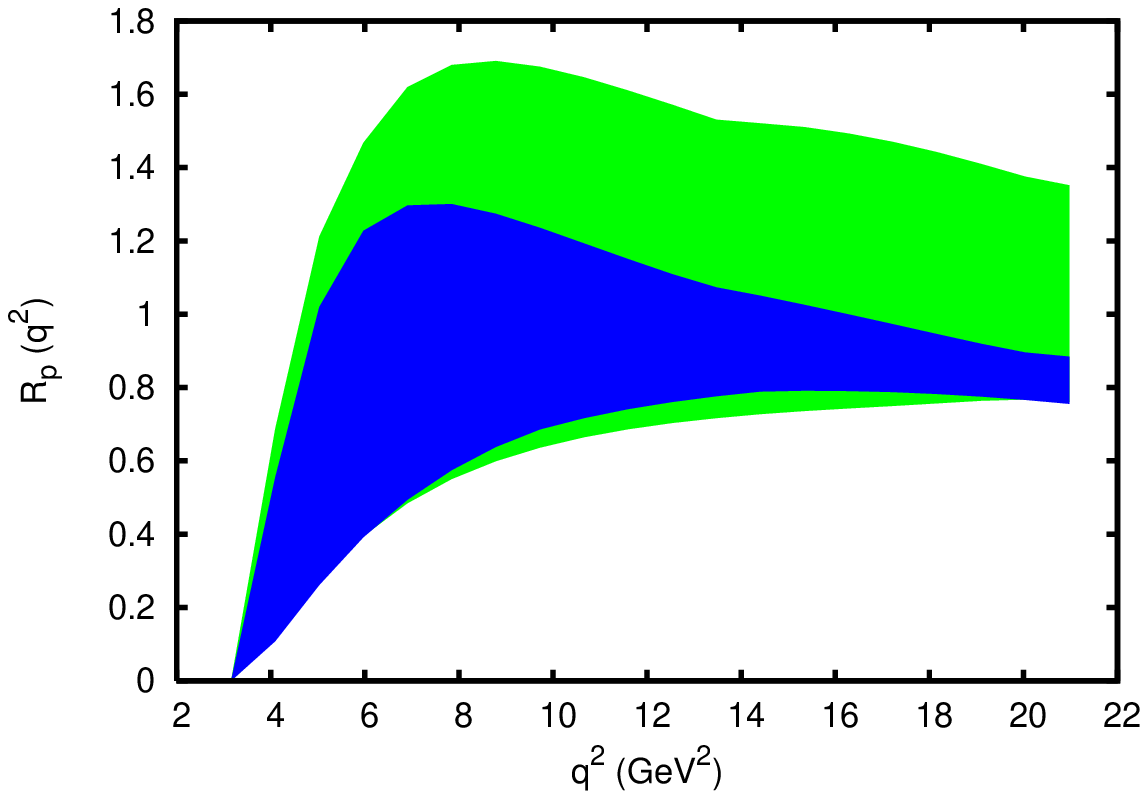}
\includegraphics[width=4cm,height=4cm]{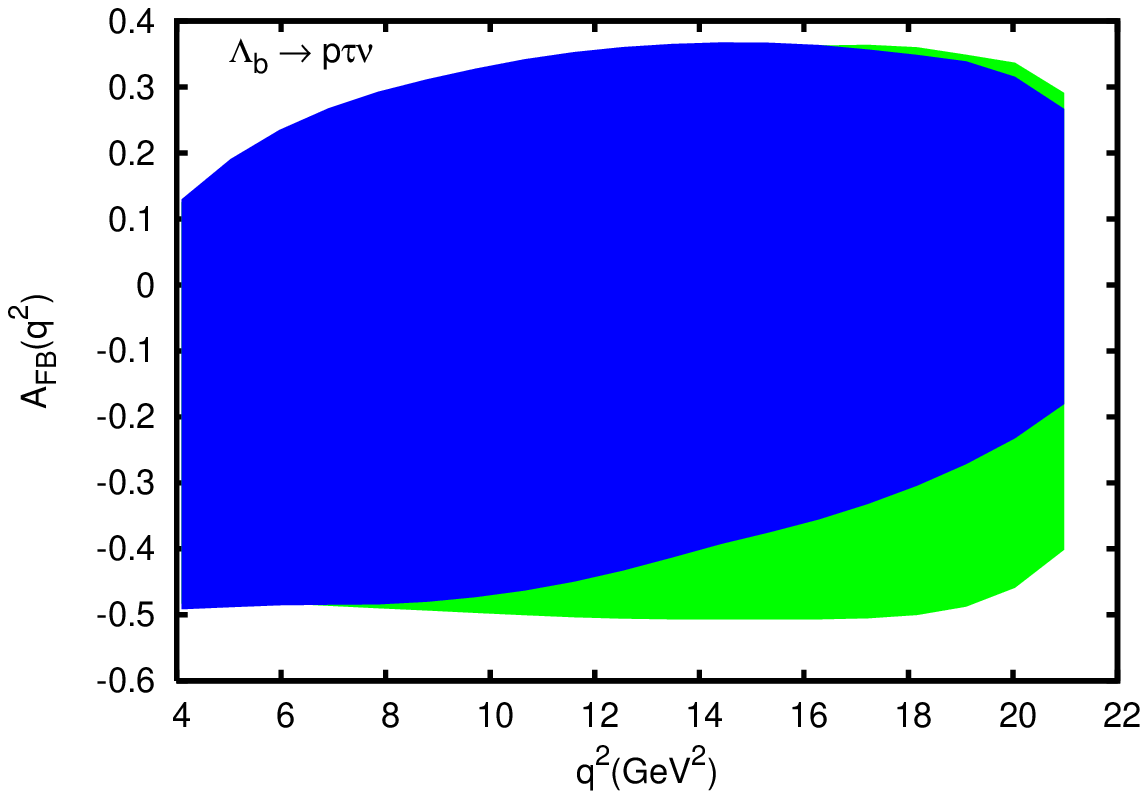}
\includegraphics[width=4cm,height=4cm]{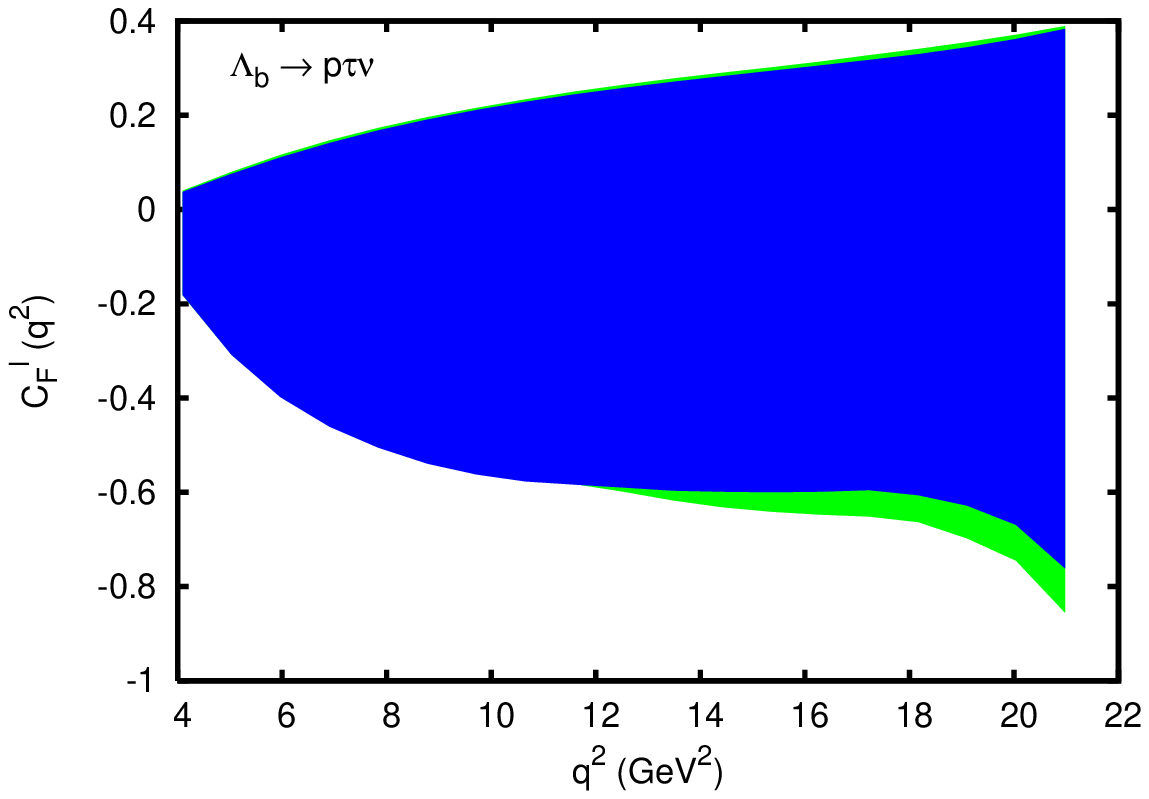}
\end{center}
\caption{The dependence of the observables DBR$(q^2)$, $R(q^2)$, $A_{FB}(q^2)$, and $C_F^l(q^2)$ on $V_L$ and $V_R$. The allowed range in
each observable is shown in light~(green) band once the NP couplings $(V_L,V_R)$ are varied within the allowed ranges of the left panel of Fig.~\ref{vlvr}.
The corresponding SM prediction is shown in dark~(blue) band. Upper and lower panel correspond to $\Lambda_b \to \Lambda_c\,\tau\,\nu$ and $\Lambda_b \to p\,\tau\,\nu$ decay modes, respectively. }
\label{obs_vlvr}
\end{figure}

Now we proceed to discuss various NP scenarios. We want to see the effect of various NP couplings in a model independent way.  In the first scenario, we assume that NP is coming from couplings associated with new vector type of interactions, i.e, from $V_L$ and $V_R$ only. We vary $V_L$ and $V_R$ while keeping $S_{L,\,R} = 0$. We impose a $3\sigma$ constraint coming from the latest experimental results on $R_D$, $R_{D^{\ast}}$, and $R_{\pi}^l$, respectively. The allowed ranges in $V_L$ and $V_R$ that satisfies the $3\sigma$ experimental constraint are shown in Fig.~\ref{vlvr}. The corresponding ranges of the branching ratios and ratio of branching ratios for the $\Lambda_b \to \Lambda_c\,\tau\nu$ and $\Lambda_b \to p\,\tau\nu$ decays are as follows:
\begin{eqnarray*}
&&\mathcal B(\Lambda_b \to \Lambda_c\,\tau\nu) = (1.51- 2.68)\times 10^{-2}\,,\qquad\qquad
\mathcal B(\Lambda_b \to p\,\tau\nu) = (1.45 - 10.92)\times 10^{-4}\,, \nonumber \\
&&R_{\Lambda_c} = (0.3213 - 0.5409)\,, \qquad\qquad
R_p = (0.5746 - 1.209)
\end{eqnarray*}
We see a significant deviation from the SM prediction. Depending on the NP couplings $V_L$ and $V_R$, value of branching ratios and the ratio of branching ratios can be either smaller or larger than the SM prediction. Precise measurement of  $\mathcal B(\Lambda_b \to \Lambda_c\,\tau\nu)$, $\mathcal B(\Lambda_b \to p\,\tau\nu)$, $R_{\Lambda_c}$, and $R_p$ will put additional constraint on the NP couplings $V_L$ and $V_R$.
We wish to look at the effect of the new physics couplings $(V_L,\,V_R)$ on different observables such as differential branching ratio ${\rm DBR}(q^2)$, ratio of branching ratio $R(q^2)$, forward backward asymmetry $A_{\rm FB}(q^2)$, and the convexity parameter $C_F^l(q^2)$ for the two decay modes. In Fig.~\ref{obs_vlvr}, we have shown in blue the allowed SM bands and in green the allowed bands of each observable once the NP couplings $V_L$ and $V_R$ are included. It can be seen that once NP is included the deviation from the SM expectation is quite large in case of ${\rm DBR}(q^2)$, $R(q^2)$, and $A_{\rm FB}(q^2)$. However, the deviation is almost negligible in case of $C_F^l(q^2)$. Again, the deviation is more in case of $\Lambda_b \to \Lambda_c\,\tau\nu$ decays compared to that of $\Lambda_b \to p\,\tau\nu$ decays. 

\begin{figure}[htbp]
\begin{center}
\includegraphics[width=8cm,height=5cm]{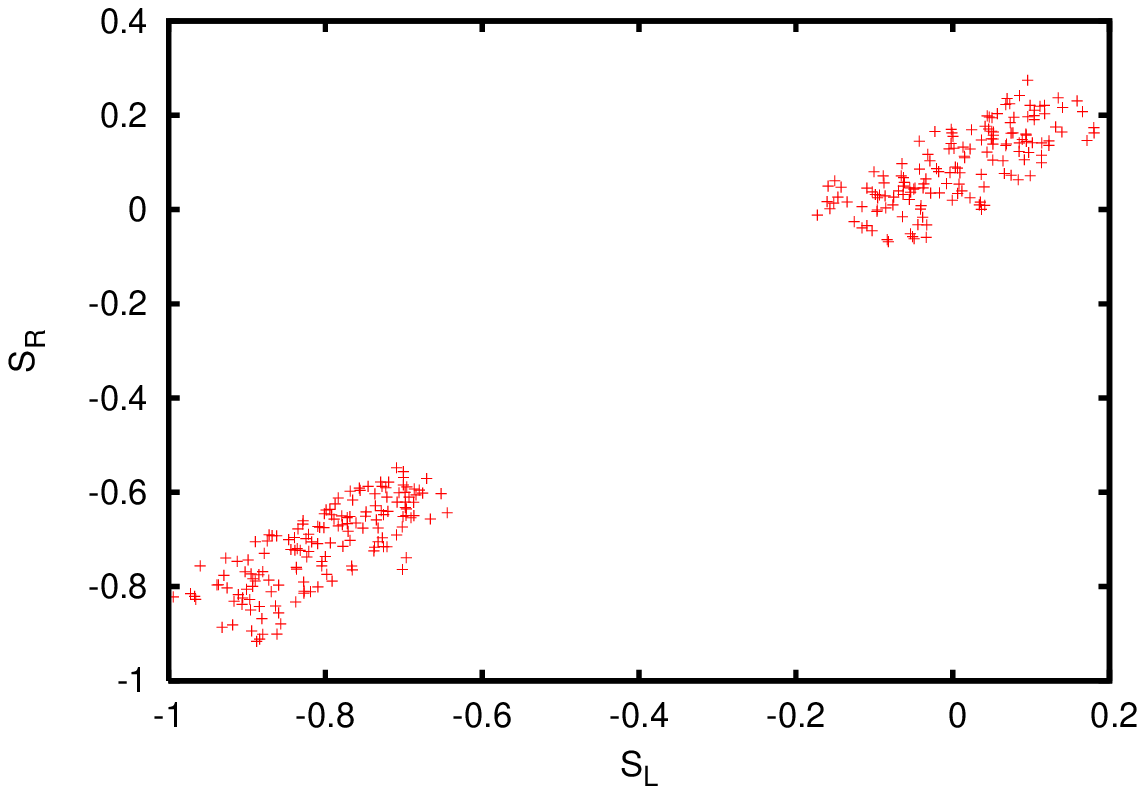}
\includegraphics[width=8cm,height=5cm]{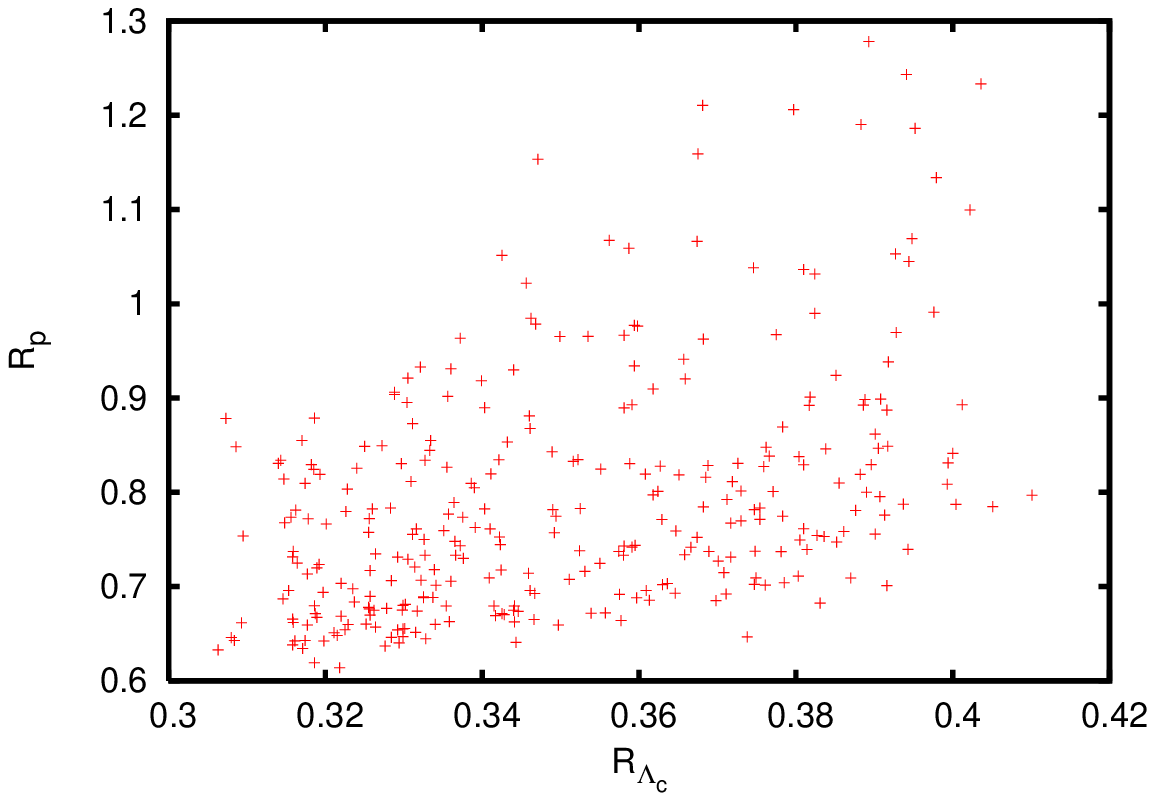}
\end{center}
\caption{Allowed regions of $S_L$ and $S_R$ obtained using the $3\sigma$ constraint coming from $R_D$, $R_{D^{\ast}}$, and $R_{\pi}^l$ are shown in the left panel and the corresponding ranges in $R_p$ and $R_{\Lambda_c}$ in
the presence of these NP couplings are shown in the right panel.}
\label{slsr}
\end{figure}

\begin{figure}[htbp]
\begin{center}
\includegraphics[width=4cm,height=4cm]{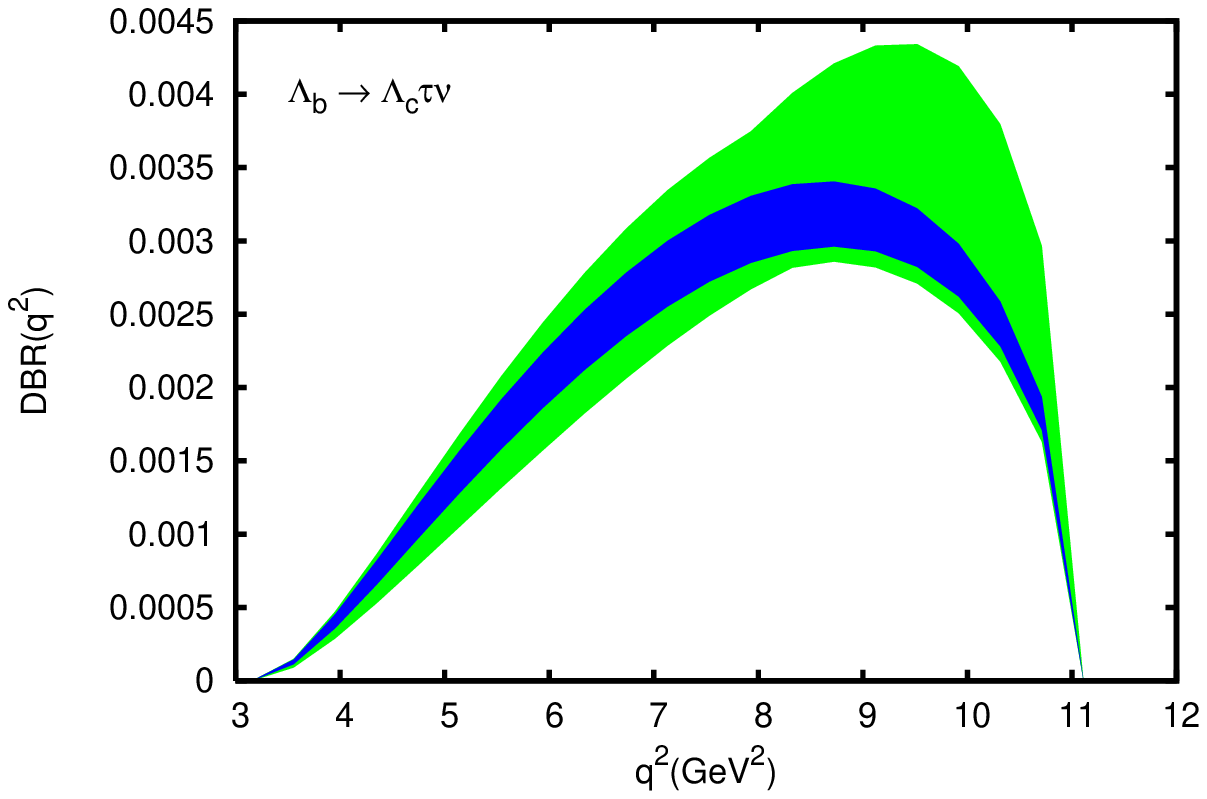}
\includegraphics[width=4cm,height=4cm]{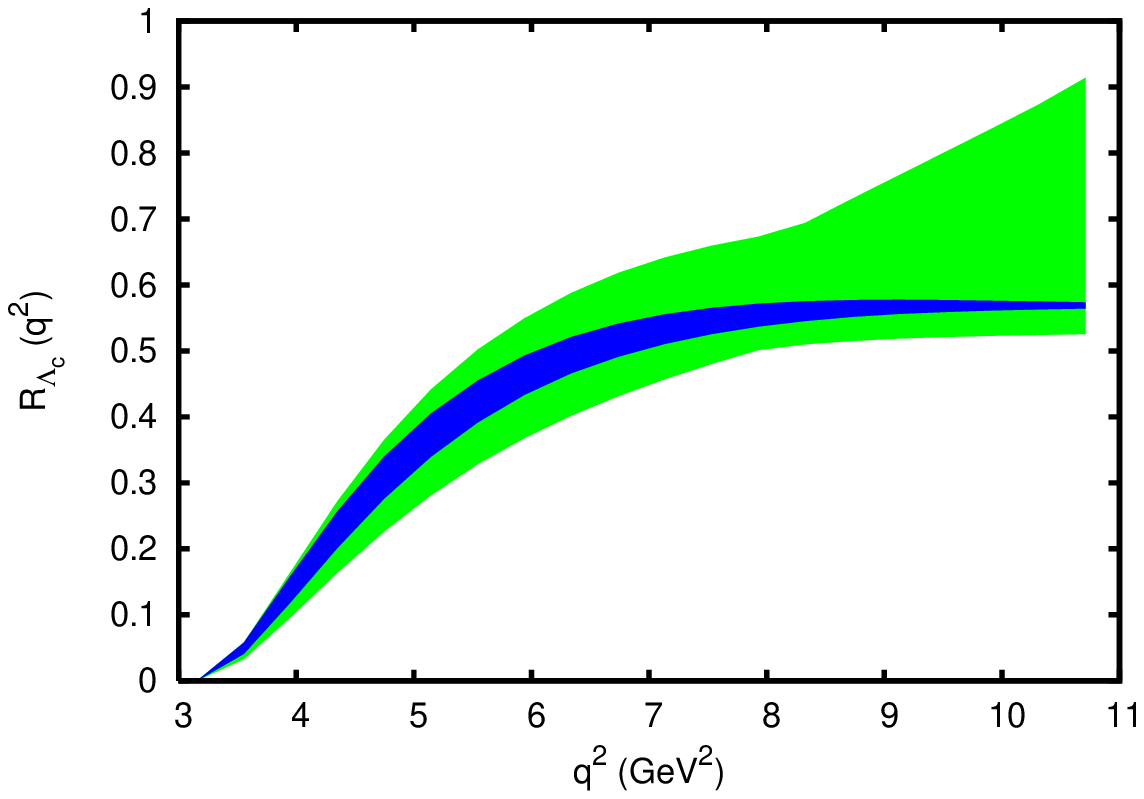}
\includegraphics[width=4cm,height=4cm]{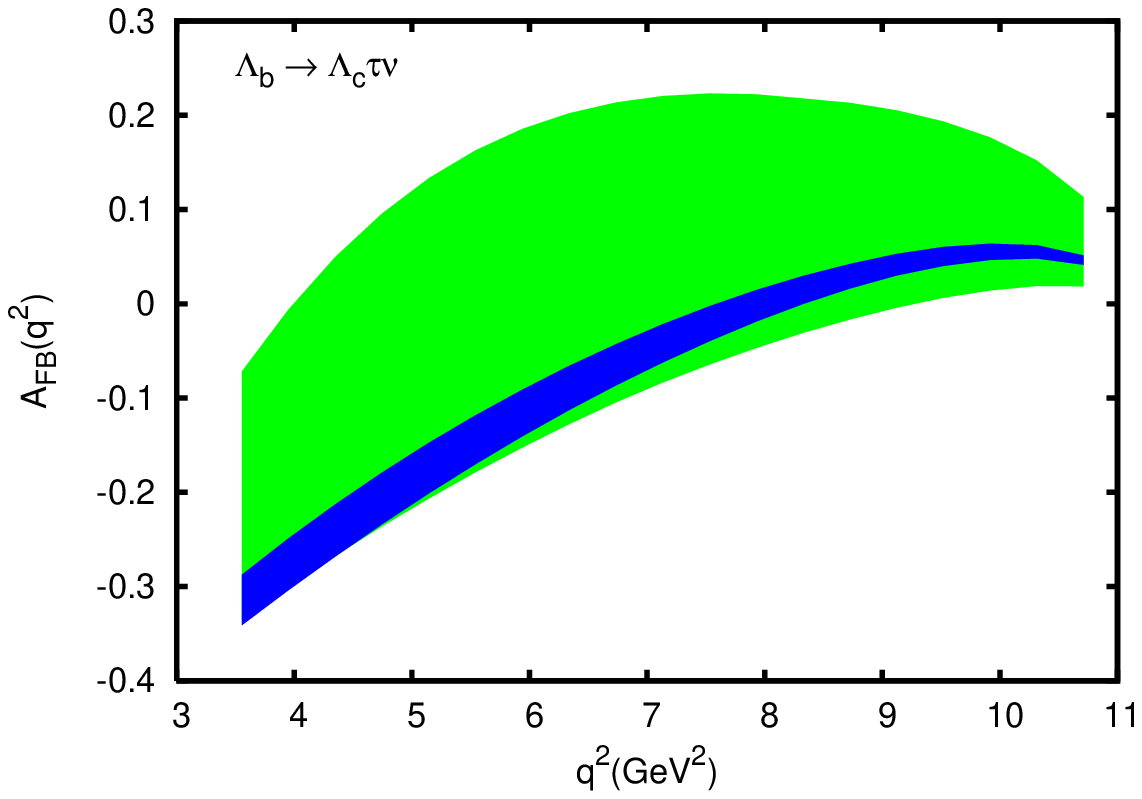}
\includegraphics[width=4cm,height=4cm]{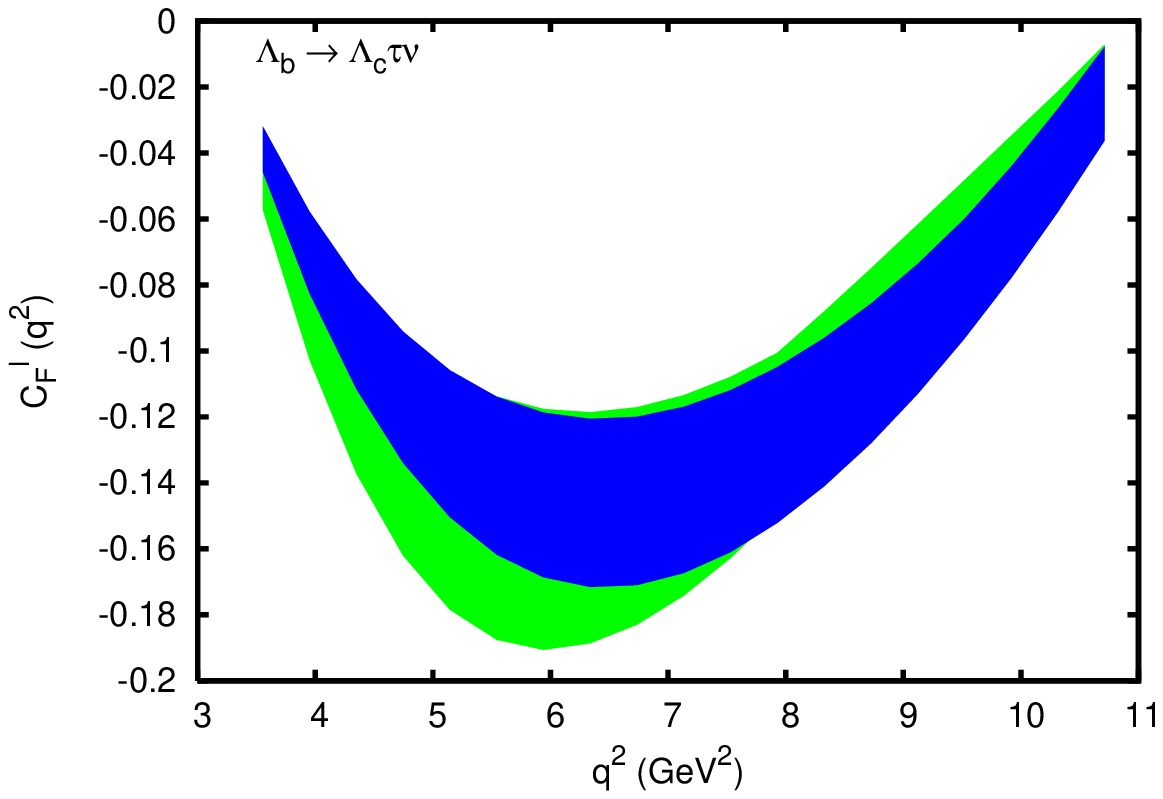}
\includegraphics[width=4cm,height=4cm]{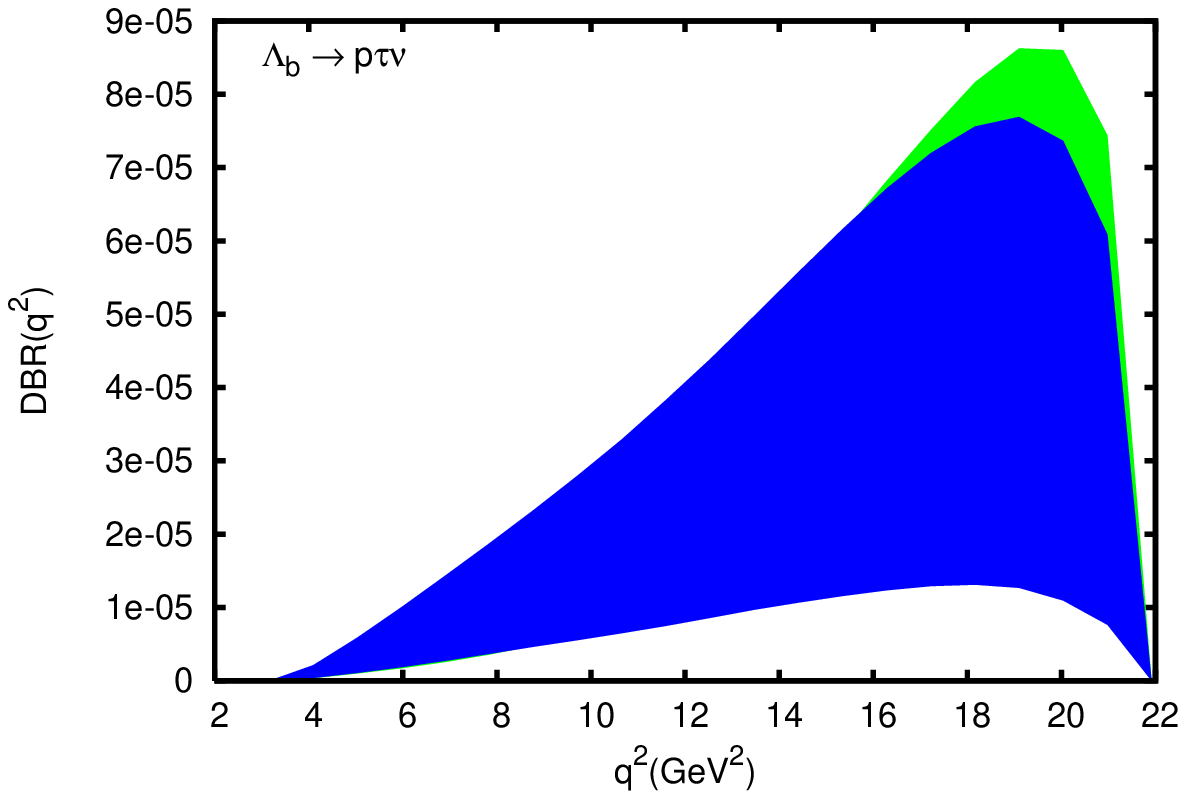}
\includegraphics[width=4cm,height=4cm]{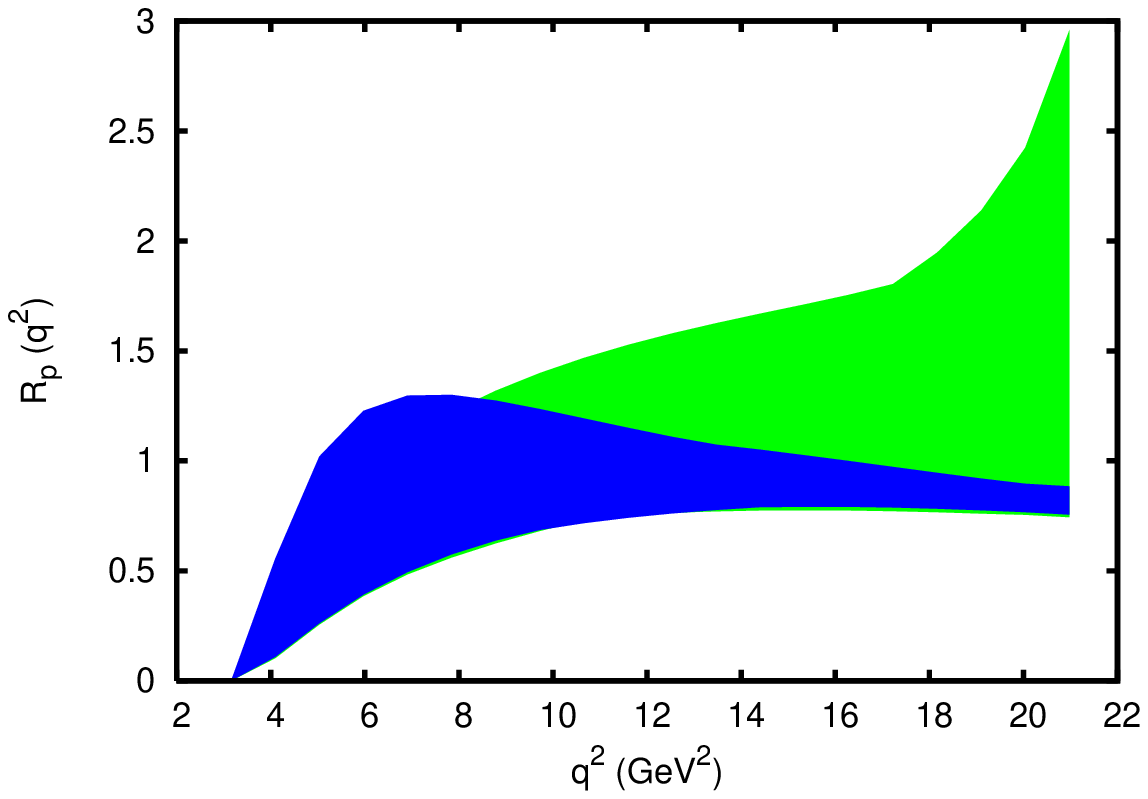}
\includegraphics[width=4cm,height=4cm]{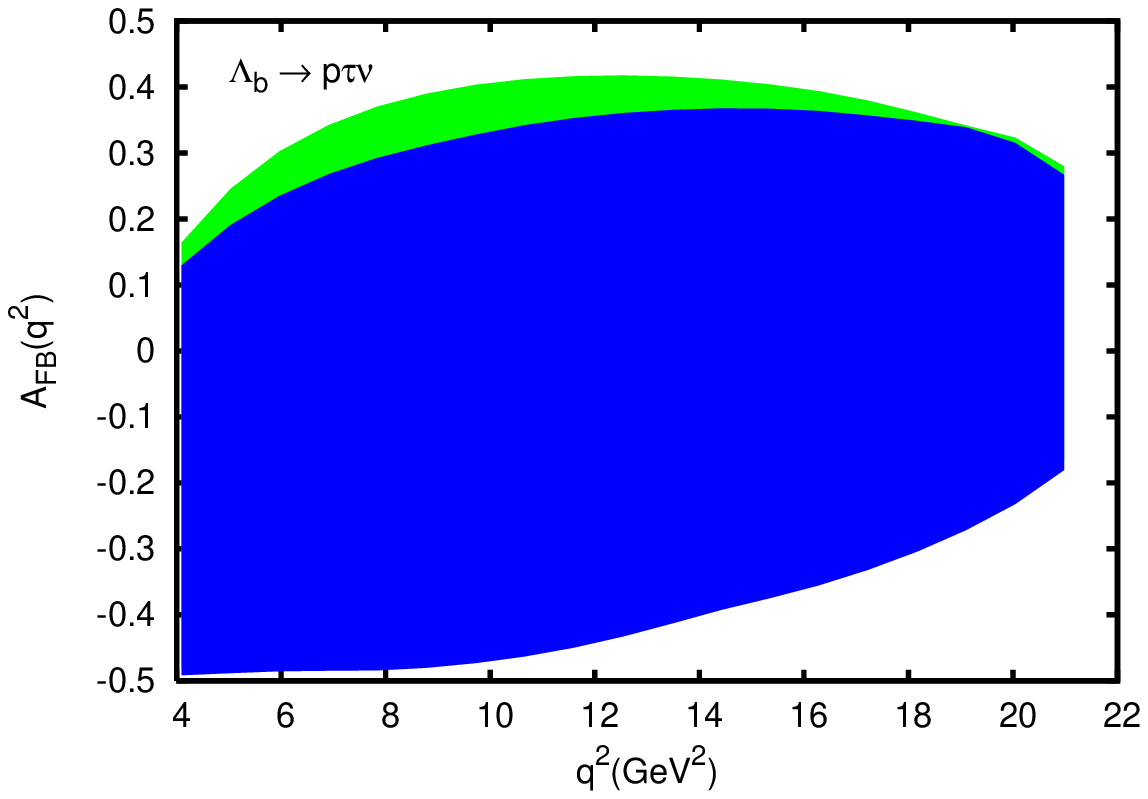}
\includegraphics[width=4cm,height=4cm]{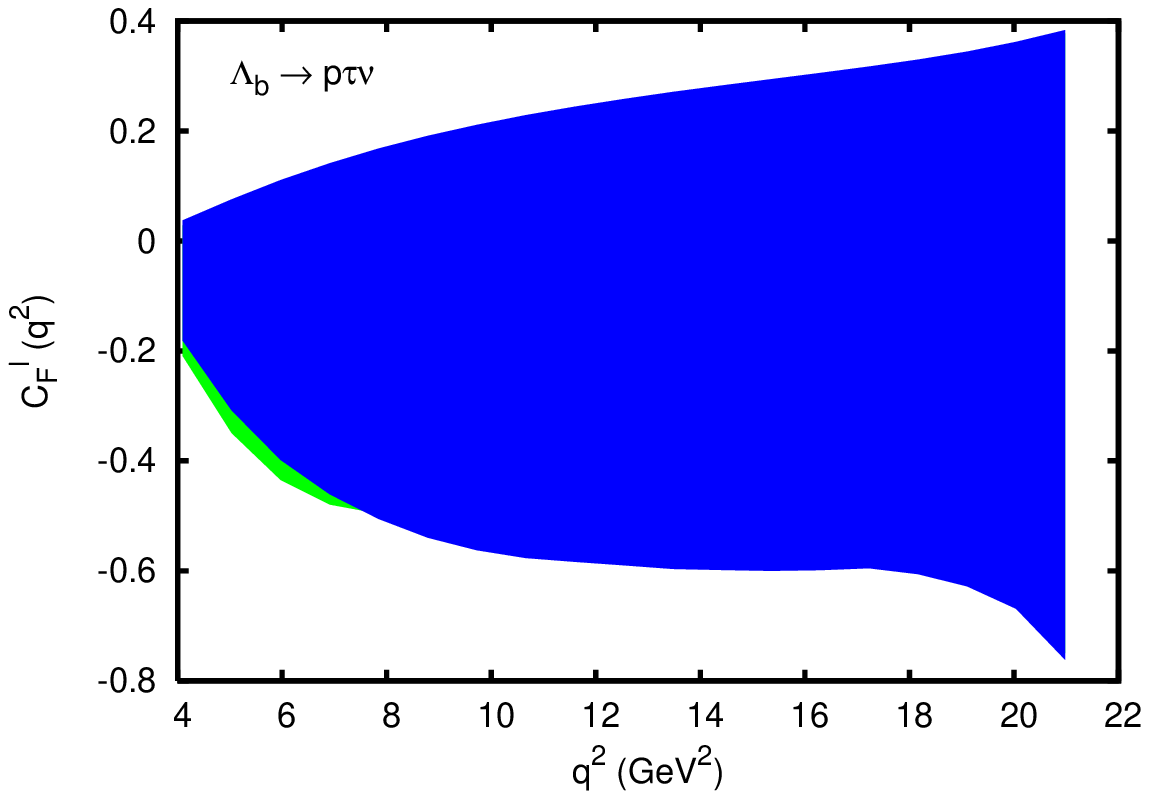}
\end{center}
\caption{The dependence of the observables DBR$(q^2)$, $R(q^2)$, $A_{FB}(q^2)$, and $C_F^l(q^2)$ on $S_L$ and $S_R$. The allowed range in
each observable is shown in light~(green) band once the NP couplings $(S_L,\,S_R)$ are varied within the allowed ranges as shown in the left panel of Fig.~\ref{slsr}.
The corresponding SM prediction is shown in dark~(blue) band. Upper and lower panel correspond to $\Lambda_b \to \Lambda_c\,\tau\,\nu$ and $\Lambda_b \to p\,\tau\,\nu$ decay modes, respectively.}
\label{obs_slsr}
\end{figure}

In the second scenario, we assume that NP is coming from new scalar type of interactions, i.e, from $S_L$ and $S_R$ only.
To explore the effect of NP coming from $S_L$ and $S_R$, we vary $S_L$ and $S_R$ and impose a $3\sigma$ constraint coming from the recent measurement of $R_D$, $R_{D^{\ast}}$, and $R_{\pi}^l$. The resulting ranges in $S_L$ and $S_R$ obtained using the $3\sigma$ experimental constraint are shown in Fig.~\ref{slsr}. In the left panel of Fig.~\ref{slsr}, the possible ranges in $R_{\Lambda_c}$ and $R_p$ are shown. The allowed ranges in all the observables are:
\begin{eqnarray*}
&&\mathcal B(\Lambda_b \to \Lambda_c\,\tau\nu) = (1.43 - 2.06)\times 10^{-2}\,,\qquad\qquad
\mathcal B(\Lambda_b \to p\,\tau\nu) = (1.60 - 7.85)\times 10^{-4}\,, \nonumber \\
&&R_{\Lambda_c} = (0.3063 - 0.4101)\,, \qquad\qquad
R_p = (0.6139 - 1.278)\,.
\end{eqnarray*}
Note that the deviation from the SM prediction can be significant depending on the values of the NP couplings $S_L$ and $S_R$.
We want to see the effect of these NP couplings on various $q^2$ dependent observables. In Fig.~\ref{obs_slsr}, we have shown how the observables ${\rm DBR}(q^2)$, $R(q^2)$, $A_{\rm FB}(q^2)$, and $C_F^l(q^2)$ behave as a function of $q^2$ with and without the NP couplings. The blue band corresponds to the SM range whereas, the green band corresponds to the NP range.
The deviations from the SM expectation is prominent in case of observables such as differential branching fraction ${\rm DBR}(q^2)$, ratio of branching fraction $R(q^2)$, and the forward backward asymmetry parameter $A_{\rm FB}(q^2)$. However, in case of the convexity parameter $C_F^l(q^2)$, the deviation is small; almost negligible for $\Lambda_b \to p\tau\nu$ decay mode. Again, it can be seen that the deviation is more pronounced in case of $\Lambda_b \to \Lambda_c\,\tau\nu$ decays compared to $\Lambda_b \to p\,\tau\nu$ decays.

We want to mention that we do not consider pure $G_V$, $G_A$, $G_S$, or $G_P$ type of NP couplings in our analysis as this kind of NP will not be able to accommodate all the existing data on $R_D$, $R_{D^{\ast}}$, and $R_{\pi}^l$ simultaneously.
\section{Conclusion}
\label{con}
Lepton flavor universality violation has been observed in various semileptonic $B$ meson decays and if it persists, it would be a definite hint of beyond the SM physics. Tensions between SM prediction  and experiments exist in various $B$ meson decays to $\tau\,\nu$ final states mediated via $b \to u$ and $b \to c$ charged current interactions such as $B \to \tau\nu$, $B \to D\,\tau\nu$, and $B \to D^{\ast}\,\tau\nu$ decays. Similar tensons have been observed in rare $B$ meson decays mediated via $b \to s\,l^+\,l^-$ transition processes as well. Recent measurement of the ratio $R_K^{\mu\,e}$ differs from SM expectation at more than $2.5\sigma$. Again, several interesting tensions between the experimental results and SM prediction have been observed in rare decays such as $B \to K^{\ast}\,\mu^+\,\mu^-$ and $B \to \phi\,\mu^+\,\mu^-$ decays. Various model dependent as well as model independent analysis have been performed in order to explain these discrepancies. Study of $\Lambda_b \to \Lambda_c\,\tau\,\nu$ and $\Lambda_b \to p\,\tau\,\nu$ decays is important mainly for two reasons. First, these decay modes are complimentary to $B \to \tau\nu$, $B \to (D,\,D^{\ast})\tau\nu$ decays mediated via $b \to c$ and $b \to u$ charged current interactions and, in principle, can provide new insights into the $R_D$, $R_{D^{\ast}}$, and $R_{\pi}^l$ puzzle. Second, precise determination of the branching fractions of these two decay modes will be useful in determining the not so well known CKM matrix elements $|V_{ub}|$ and $|V_{cb}|$. 

We study $\Lambda_b \to \Lambda_c\,l\,\nu$ and $\Lambda_b \to p\,l\,\nu$ decays mediated via $b \to u$ and $b \to c$ transitions within the context of an effective Lagrangian in the presence of NP. Similar approach was also adopted in Ref.~\cite{Shivashankara:2015cta}. However, in our work, we consider both $\Lambda_b \to \Lambda_c\,l\,\nu$, and $\Lambda_b \to p\,l\,\nu$ decays, mediated via $b \to u$ and $b \to c$ charged current interactions, within one framework and perform a combined analysis using the $3\sigma$ constraint coming from the most recent experimental results on $R_D$, $R_{D^{\ast}}$, and $R_{\pi}^l$  to explore the pattern of NP. This is where we differ significantly from Ref.~\cite{Shivashankara:2015cta}. Moreover, the various $\Lambda_b \to \Lambda_c$ and $\Lambda_b \to p$ transition form factors that we use are also different from Ref.~\cite{Shivashankara:2015cta}. We assume NP in the third generation leptons only. We also assume the NP couplings to be real for our analysis. We look at two different NP scenarios.  Now let us summarize our main results.

We first report the central values and the $1\sigma$ ranges in the branching fractions, the ratio of branching fractions, and the ratio of partially integrated decay rates of $\Lambda_b \to \Lambda_c\,l\,\nu$ and $\Lambda_b \to p\,l\,\nu$ decay modes within the SM. Our value of $R_{\Lambda_c}$ is exactly same as in Ref.~\cite{Detmold:2015aaa}, however, it differs slightly from the value reported in Refs.~\cite{Woloshyn:2014hka, Shivashankara:2015cta, Gutsche:2015mxa}. It is due to the fact that we use the latest lattice calculations of the form factors from Ref.~\cite{Detmold:2015aaa}.

We include vector and scalar type of NP interactions in our analysis and explore two different NP scenarios.
In the first scenario, we consider only vector type of NP interactions, i.e, we consider that only $V_L$ and $V_R$ contributes to these two decays modes. We find the possible ranges in $V_L$ and $V_R$ using the $3\sigma$ constraint coming from the most recent experimental results on $R_D$, $R_{D^{\ast}}$, and $R_{\pi}^l$. The range in $R_{\Lambda_c}$ and $R_p$ with these NP couplings are found to be $[0.3213, 0.5409]$ and $[0.5746, 1.209]$, respectively. We also study the dependence of various $q^2$ dependent observables such as ${\rm DBR}(q^2)$, $R(q^2)$, $A_{\rm FB}(q^2)$, and $C_F^l(q^2)$ on the NP parameters $V_L$ and $V_R$. We find significant deviations from the SM prediction once the NP couplings are included. However, the deviation from the SM prediction is more pronounced in case of $\Lambda_b \to \Lambda_c\,\tau\,\nu$ decay mode. 

 In the second NP scenario, we assume that NP is coming only from scalar type of interactions, i.e, from $S_L$ and $S_R$ only. We use $3\sigma$ experimental constraint coming from the recent measurement of $R_D$, $R_{D^{\ast}}$, and $R_{\pi}^l$ to find the allowed ranges in $S_L$ and $S_R$. The range in $R_{\Lambda_c}$ and $R_p$ with these NP couplings are found to be $[0.3063, 0.4101]$ and $[0.6139, 1.278]$, respectively. It is noted that the parameter space is somewhat more constrained in this scenario. Again, for the $q^2$ dependent observables, we see significant deviations from the SM predictions in all the observables. Similar to the first scenario, we see that the deviation from the SM prediction is more pronounced in case of $\Lambda_b \to \Lambda_c\,\tau\,\nu$ decay mode.
 
 Although, there is hint of NP in various leptonic and semileptonic $B$ decays, NP is not yet established. Reduced theoretical uncertainties in the hadronic form factors, decay constants, and the CKM matrix elements will certainly help in disentangling the NP from the SM uncertainties. Again,  more precise measurements are also needed to confirm the presence of NP.  Measurement of all the observables for $\Lambda_b \to \Lambda_c\,\tau\nu$ and $\Lambda_b \to p\,\tau\nu$ decay modes will be crucial to test for various NP patterns. At  the same time, precise determination of $\Lambda_b \to \Lambda_c$ and $\Lambda_b \to p$ transition form factors will also help in determining the poorly known CKM matrix element $|V_{ub}|$.

\end{document}